\documentclass[peerreview]{IEEEtran}

\usepackage{amsthm}
\usepackage{amsmath}
\usepackage{mathrsfs}
\usepackage{amssymb} 
\usepackage{bbm,dsfont} 
\usepackage{multirow} 
\usepackage{units}
\usepackage{stmaryrd}   
\usepackage{verbatim}
\usepackage{enumerate} 
\usepackage[table]{xcolor}
\usepackage{slashbox} %
\usepackage{ wasysym }
\usepackage{tikz}
\usetikzlibrary{calc}
\usepackage{tabularx}

\title{Communication over Quantum Channels with Parameter Estimation}
\author{\IEEEauthorblockN{Uzi Pereg}\\
\IEEEauthorblockA{Institute for Communications Engineering, \\
Technical University of Munich,
 80333 Munich, Germany.\\
Email: {\tt uzi.pereg@tum.de}
 }}

\usepackage[square,numbers,sort&compress]{natbib}
\usepackage{xcolor}
\usepackage{hyperref} %
\usepackage%
{hyperref} %
\hypersetup{
    colorlinks,
    linkcolor={blue!80!black},%
    citecolor={green!50!black},
    urlcolor={blue!80!black}
}

\usepackage{graphicx}
\graphicspath{{images/}}

\newcommand\blfootnote[1]{%
  \begingroup
  \renewcommand\thefootnote{}\footnote{#1}%
  \addtocounter{footnote}{-1}%
  \endgroup
}

\newcommand{\bieee}{\begin{IEEEeqnarray}{rCl}}
\newcommand{\eieee}{\end{IEEEeqnarray}}
\newcommand{\prob}[1]{\Pr\left(#1\right)}
\newcommand{\given}{\mid}
\newcommand{\cprob}[2]{\Pr\left(#1\given #2\right)}
\newcommand{\E}{\mathbb{E}}

\renewcommand{\mathbbm}[1]{\text{\usefont{U}{bbm}{m}{n}#1}} %
\newcommand{\eps}{\varepsilon}

\newcommand{\norm}[1]{\left\lVert#1\right\rVert}
\newcommand{\trace}{\mathrm{Tr}}

\newcommand{\identity}{\mathbbm{1}}
\newcommand{\kb}[1]{ | #1 \rangle\langle #1 | } %

\newcommand{\ie}{\emph{i.e.} }
\newcommand{\eg}{\emph{e.g.} }
\newcommand{\etal}{\emph{et al.} }
\newcommand{\cf}{\emph{cf.} }

\newcommand{\tS}{\widetilde{S}}

\newcommand{\tR}{\widetilde{R}}

\newcommand{\ha}{\hat{a}}
\newcommand{\hb}{\hat{b}}

\newcommand{\he}{\hat{e}}

\newcommand{\hm}{\hat{m}}
\newcommand{\hs}{\hat{s}}

\newcommand{\hK}{\hat{K}}
\newcommand{\hP}{\hat{P}}
\newcommand{\hM}{\hat{M}}
\newcommand{\hL}{\hat{L}}
\newcommand{\hS}{\hat{S}}

\newcommand{\Aset}{\mathcal{A}}
\newcommand{\Bset}{\mathcal{B}}

\newcommand{\Fset}{\mathcal{F}}

\newcommand{\Gset}{\mathcal{G}}
\newcommand{\Hset}{\mathcal{H}}

\newcommand{\Kset}{\mathcal{K}}

\newcommand{\Mset}{\mathcal{M}}
\newcommand{\Pset}{\mathcal{P}}

\newcommand{\Vset}{\mathcal{V}}

\newcommand{\Sset}{\mathcal{S}}
\newcommand{\Wset}{\mathcal{W}}
\newcommand{\Xset}{\mathcal{X}}
\newcommand{\Yset}{\mathcal{Y}}
\newcommand{\Zset}{\mathcal{Z}}
\newcommand{\Eset}{\mathcal{E}}
\newcommand{\markovC}[1]{%
\begin{tikzpicture}[#1]%
\draw (0,0.3ex) -- (1ex,0.3ex);%
\draw (0.5ex,0.3ex) circle (0.2ex);
\draw[white] (0.2ex,0) -- (0.5ex,0);%
\end{tikzpicture}%
}
\newcommand{\Cbar}{\markovC{scale=2}}

\theoremstyle{remark}	\newtheorem{theorem}{Theorem}
\theoremstyle{remark}	\newtheorem{lemma}[theorem]{Lemma}
\theoremstyle{remark}	\newtheorem{corollary}[theorem]{Corollary}
\theoremstyle{remark}	
\theoremstyle{remark} \newtheorem{definition}{Definition}
\theoremstyle{remark} \newtheorem{remark}{Remark}
\theoremstyle{remark} \newtheorem{example}{Example}

\newcommand{\pSpace}{\mathcal{P}}														%

\newcommand{\tset}{\Aset^{\delta}}													%
\newcommand{\Tset}{\mathcal{T}}												%

\newcommand{\channel}{\mathcal{N}}

\newcommand{\inC}{\mathsf{C}}
\newcommand{\inR}{\mathsf{R}}
\newcommand{\opC}{\mathbb{C}}

\begin{document}
\maketitle

{}

\begin{abstract}
Communication over a random-parameter quantum channel when the decoder is required to reconstruct the parameter sequence is considered.
We study scenarios that include either strictly-causal, causal, or non-causal channel side information (CSI) available at the encoder, and also when CSI is not available. 
This model can be viewed as a form of quantum metrology, and as the quantum counterpart of the classical rate-and-state channel with state estimation at the decoder.
Regularized formulas for the capacity-distortion regions are derived. 
In the special case of measurement channels, single-letter characterizations are derived for the strictly-causal and causal settings.
Furthermore, in the more general case of entanglement-breaking channels, a single-letter characterization is derived when CSI is not available.
As a consequence, we obtain regularized formulas for the capacity of random-parameter  quantum channels with CSI, %
generalizing previous results by Boche et al. \cite{BocheCaiNotzel:16p} on classical-quantum channels.
Bosonic dirty paper coding is introduced as a consequence, %
where we demonstrate that the optimal coefficient %
is not necessarily that of minimum mean-square error estimation as in the classical setting.
\end{abstract}

\begin{IEEEkeywords}
Quantum communication, Shannon theory, state estimation, rate-and-state channel, bosonic channel, writing on dirty paper, encoding constraints.
\end{IEEEkeywords}

\blfootnote{
Parts of this work have been presented  at the 2020 IEEE International
Symposium on Information Theory, Los Angeles, CA, USA, June 21 – 26, 2020, and the 2021 IEEE International
Symposium on Information Theory, Melbourne, Victoria, Australia, July 12 – 20, 2021.
}

\section{Introduction}
\label{intro}
A fundamental task in classical information theory is to determine the ultimate transmission rate of communication.
Various %
settings of practical significance can be described by a channel $p_{Y|X,S}$ that depends on a random parameter $S\sim q(s)$ when there is channel side information (CSI) available at the transmitter %
\cite{%
KeshetSteinbergMerhav:07n,ChoudhuriKimMitra:13p,PeregSteinberg:19p1}. %
For example,    a cognitive radio in a wireless system may be aware of the channel state and network configuration \cite{GJMS:09p}. %
Other applications include 
memory storage where the writer knows the fault locations \cite{HeegardElGamal:83p}, %
 digital watermarking \cite{MoulinOsullivan:03p}, and spread-spectrum communication  \cite{ChenWornell:01p,SedghiKhademiCvejic:06c},  %
 where the CSI represents the host data or a pseudo-random sequence to be modulated.

In the rate-and-state (RnS) model \cite{SutivongChiangCoverKim:05p}, the receiver is not only required to recover the message, but also to estimate the parameter sequence with limited distortion. 
For example, in
digital multicast \cite{SutivongChiangCoverKim:05p}, %
the message represents digital control information that is multicast on top of an existing analog transmission, which is also estimated by the receiver. Additional applications can be found in \cite{ChoudhuriKimMitra:13p} and references therein. 
 The capacity-distortion tradeoff region with strictly-causal CSI and with causal CSI was determined by Choudhuri \etal \cite{ChoudhuriKimMitra:13p}, and without CSI by Zhang \etal \cite{ZhangVedantamMitra:08c,ZhangVedantamMitra:11p}. Inner and Outer bounds on the tradeoff region with non-causal CSI were derived by Sutivong in \cite{Sutivong:03z}, with full characterization in the Gaussian case \cite{SutivongChiangCoverKim:05p}.
The RnS channel with feedback was recently considered by Bross and Lapidoth \cite{BrossLapidoth:18p}.

The field of quantum information is rapidly evolving in both practice and theory 
\cite{DowlingMilburn:03p,JKLGD:13p,OrieuxDiamanti:16p,FlaminiSpagnoloSciarrino:18r,PERLHPCVV:20p%
}.
Quantum information theory is the natural extension of classical information theory. Nevertheless, 
this generalization reveals astonishing phenomena with no parallel in classical communication \cite{GyongyosiImreNguyen:18p}. For example, 
 two quantum channels, each with zero quantum capacity, can have a nonzero quantum capacity
when used together \cite{SmithYard:08p}. This property is known as super-activation.

Communication through quantum channels can be separated into different categories.
The Holevo-Schumacher-Westmoreland (HSW) Theorem provides a regularized (``multi-letter")  formula for the capacity of a quantum channel \cite{Holevo:98p,SchumacherWestmoreland:97p}. %
Although calculation of such a formula is intractable in general, it provides computable lower bounds, and there are special cases where the capacity can be computed exactly. The reason for this difficulty is that the Holevo information is not necessarily additive \cite{Hastings:09p,Holevo:12b}. %
Shor has demonstrated additivity %
for the class of entanglement-breaking channels  \cite{Shor:02p}, in which case the HSW theorem provides a single-letter computable formula for the capacity. This class includes both classical-quantum channels and 
measurement (quantum-classical) channels \cite[Section 4.6.7]{Wilde:17b}.
A similar difficulty occurs with transmission of quantum information
\cite{%
Devetak:05p}.

As for quantum channels with random parameters, Boche, Cai, and N\"{o}tzel \cite{BocheCaiNotzel:16p} addressed the classical-quantum channel  with CSI at the encoder. The capacity was determined given causal CSI, and a regularized formula was provided given %
non-causal CSI. Warsi and Coon \cite{WarsiCoon:17p} used an information-spectrum approach to derive  multi-letter bounds for a similar setting, where the side information has a limited rate. 
Anshu \etal \cite{AnshuHayashiWarsi:20p} have recently considered
the fully quantum wiretap channel with CSI as well.
The entanglement-assisted capacity of a quantum channel with non-causal CSI was determined by    Dupuis \cite{Dupuis:08a,Dupuis:09c}, and with causal CSI by the author \cite{Pereg:19c3,Pereg:19a}. One-shot communication with  CSI is considered in \cite{AnshuJainWarsi:19p} as well.
 Luo and Devetak \cite{LuoDevetak:09p} considered channel simulation with source side information (SSI) at the 
decoder, and also solved the quantum generalization of the Wyner-Ziv problem \cite{WynerZiv:76p}.  Quantum data  compression with SSI  is also studied in \cite{%
DattaHircheWinter:19c,%
CHDH:19c,KhanianWinter:20p}. %
State-dependent channels with 
environment assistance are considered in \cite{Smolin:05p,Winter:05a,OskoueiManciniWinter:21a}. %
The dual setting of state masking, where the channel state is hidden from the receiver, was recently considered in 
\cite{PeregDeppeBoche:21p}.
Quantum relay channels are treated in \cite{SavovWildeVu:12c,DGHW:19p} using a decode-forward communication scheme with block Markov coding. 
Parameter estimation of quantum channels was previously studied from the algorithmic point of view in different settings \cite{Fujiwara:05p,JWDFY:08p,ZorziTicozziFerrante:11a}.

\begin{center}
\begin{figure}[tb]
\includegraphics[scale=0.7,trim={-5.5cm 12cm 3cm 12.25cm},clip]
{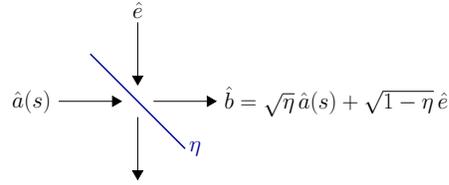} %
\caption{The beam-splitter relation of the single-mode bosonic channel. The channel input is an electromagnetic field mode with an annihilation operator $\ha$, and the output is another mode with the annihilation operator $\hb=\sqrt{\eta}\, \ha +\sqrt{1-\eta}\,\he$,
where $\he$ is associated with the environment noise and the parameter $\eta$  is %
the transmissivity, where $0\leq \eta\leq 1$.
}
\label{fig:BSp}
\end{figure}

\end{center}

Optical communication forms %
the backbone of the Internet \cite{BardhanShapiro:16p,Savov:12z,KumarDeen:14b}.
The bosonic channel is a simple quantum-mechanical model for optical communication over free space or optical fibers 
\cite{WPGCRSL:12p,WildeHaydenGuha:12p}, and it can be viewed as the quantum counterpart of the classical channel with additive white Gaussian noise (AWGN).
 An optical communication system consists of a modulated source of photons, the optical channel, and an optical detector. %
For a single-mode bosonic channel, the channel input is an electromagnetic field mode with an annihilation operator $\ha$, and the output is another mode with the annihilation operator $\hb$. The 
input-output relation in the Heisenberg picture \cite{HolevoWerner:01p} is given by %
\begin{align}
\hb=\sqrt{\eta}\, \ha +\sqrt{1-\eta}\,\he
\end{align}
where $\he$ is associated with the environment noise and the parameter $\eta$  is %
the transmissivity, $0\leq \eta\leq 1$, which depends on the length of the optical fiber and its absorption length \cite{EisertWolf:05c} (see Figure~\ref{fig:BSp}). 
For a lossy bosonic channel, the noise mode $\he$ is in a Gibbs thermal state. 
 Modulation is performed such that the unitary displacement operator $D(\alpha)=\exp(\alpha \ha^\dagger-\alpha^* \ha)$ is applied to the vacuum state $\kb{0}$ \cite{WPGCRSL:12p}. 

\vspace{-0.75cm}

In this paper, we consider a random-parameter quantum channel when the decoder is required to reconstruct the parameter sequence in a lossy manner, \ie with limited distortion. 
Here, we give two applications for this model: digital multicast using \emph{quantum} communication channels, and classical watermarking with a quantum embedding. 
 In the watermarking application, an authentication message is mixed within
 classical host data (``stegotext''),  and this mixture is encoded into a quantum state that is sent to an authenticator. The random parameters in this setting represent the host data, while their
 estimation at the decoder corresponds to a scenario where
 the host data itself contains desirable information.
Our setting can also be interpreted as a form of quantum metrology \cite{GiovannettiLloydMaccone:11p}, where the decoder performs  measurements on the received (quantum) systems in order to estimate classical noise parameters, while exploiting the entanglement generated by the encoder.

The scenarios that are studied in the present work include either strictly-causal, causal, or non-causal channel side information (CSI) available at the encoder, as well as the case where CSI is not available. With strictly-causal CSI, Alice knows the \emph{past} parameters at each time instance; given causal CSI, she knows the \emph{past and present} parameters; and with non-causal CSI, the entire sequence of random parameters is available to her a priori.
 This model can be viewed as the quantum analog of the classical RnS channel.
We derive regularized formulas for the capacity-distortion tradeoff regions. %
In the special case of measurement channels,  single-letter characterizations are established for the strictly-causal and causal settings. %
Furthermore, in the more general case of entanglement-breaking channels, a single-letter characterization is derived when CSI is not available.
 As a consequence, we obtain regularized formulas for the capacity of random-parameter  quantum channels with strictly-causal, causal, or non-causal CSI, %
generalizing the previous results by Boche \etal \cite{BocheCaiNotzel:16p} on classical-quantum channels.

Considering entanglement-breaking channels without CSI, we use a different approach from that of Shor \cite{Shor:02p}.
As opposed to Shor \cite{Shor:02p}, we do not show additivity of the capacity formula, but rather extend the methods of Wang \etal
\cite{WangDasWilde:17p} to prove the converse part in a more direct manner. 
To prove achievability with strictly-causal CSI, we extend  the classical block Markov coding method from \cite{ChoudhuriKimMitra:13p} to the quantum setting, and then apply the quantum packing lemma \cite{HsiehDevetakWinter:08p} for decoding the message, and the classical covering lemma for the reconstruction of the parameter sequence.
The gentle measurement lemma \cite{Winter:99p,OgawaNagaoka:07p} alleviates the proof, as it guarantees that multiple decoding measurements can be performed without collapsing the quantum state and such that the output state after each measurement is almost the same. Thus, we can separate between measurements for recovering the message and for sequence reconstruction.
Achievability with causal CSI is proved using similar techniques with the addition of a quantum ``Shannon-strategy" encoding operation  \cite{Shannon:58p} %
\cite[Section IV.D]{Pereg:19a}. %
To prove achievability with non-causal CSI, we use an extension of  the classical binning technique \cite{HeegardElGamal:83p} to the quantum setting. %

Furthermore, we introduce the bosonic dirty paper setting as a special case.  We consider the single-mode lossy bosonic channel with a coherent-state protocol and a non-ideal displacement operation in the modulation process:
\begin{align}
|\zeta_1 \zeta_2 \cdots \zeta_n \rangle =D(\alpha_1+s_1)|0\rangle %
\otimes \cdots \otimes
D(\alpha_n+s_n)|0\rangle
\label{eq:Zetan}
\end{align}
where the parameter $s_i$ represents classical interference in the transmission equipment, which the transmitter becomes aware of, while the receiver is not. It is assumed that the input has %
an average power constraint $\frac{1}{n}\sum_{i=1}^n |\alpha_i|^2 \leq N_A$.  %
Alternatively, this can be viewed as a watermarking model with a quantum embedding.
Given a classical host data sequence $s_1,\ldots,s_n$, %
Alice encodes an authentication message $m$ into a watermark $(\alpha_i(m,s_1,\ldots,s_n))_{i=1}^n$. Next, Alice performs a quantum embedding of the watermark; 
she prepares a \emph{watermarked state} $|\zeta_1 \zeta_2 \cdots \zeta_n \rangle$ as in (\ref{eq:Zetan}), %
and transmits it to the authenticator Bob through the optical fiber.
The capacity of the random-parameter bosonic channel represents the optimal rate at which the authenticator can recover the messages with high fidelity. %

First, we consider homodyne and heterodyne detection. Both of those settings reduce to a classical random-parameter channel with either real or complex-valued Gaussian noise. Thereby, we observe that based on Costa's dirty-paper solution, the effect of the classical interference can be canceled, and the capacity is the same regardless of the intensity of the interference. 
Then, we consider joint detection, in which case, the problem does not reduce to that of a classical description.
We derive a dirty-paper coding lower bound based on the above results, with a general coefficient $t$ (see (\ref{eq:UtX})).
Considering the special case of a pure-loss bosonic channel, we show that the optimal coefficient 
is not necessarily that of minimum mean-square error (MMSE) estimation  value as in (\ref{eq:tMMSE}).

The paper is organized as follows. In Section~\ref{sec:definitions}, we give the definitions and present the models. In Section~\ref{subsec:Previous}, we provide a brief review of related work on channels without random parameters, regularization, additivity, and entanglement-breaking channels; as well as a 
comparison between Shor's original approach for single-letterization, based on additivity, and the alternative argument that extends the methods in 
\cite{WangDasWilde:17p}.
In Section~\ref{sec:main},
we state our main results on the random-parameter quantum channel with parameter estimation at the decoder. %
In Section~\ref{sec:bosonic}, we consider bosonic dirty paper coding as a consequence of the main results.
Section~\ref{sec:discussion} is dedicated to summary and discussion, where we summarize our main results and conclude with remarks on the comparison between the classical and quantum dirty-paper settings. 
The proofs are given in the appendix.

\section{Definitions}
\label{sec:definitions}
\subsection{Notation, States, and Information Measures}
 We use the following notation conventions. %
Calligraphic letters $\Xset,\Yset,\Zset,...$ are used for finite sets.
Lowercase letters $x,y,z,\ldots$  represent constants and values of classical random variables, and uppercase letters $X,Y,Z,\ldots$ represent classical random variables.  
 The distribution of a  random variable $X$ is specified by a probability mass function (pmf) 
	$p_X(x)$ over a finite set $\Xset$. %
 We use $x^j=(x_1,x_{2},\ldots,x_j)$ to denote  a sequence of letters from $\Xset$. %
 A random sequence $X^n$ and its distribution $p_{X^n}(x^n)$ are defined accordingly. 
For a pair of integers $i$ and $j$, $1\leq i\leq j$, we write a discrete interval as $[i:j]=\{i,i+1,\ldots,j \}$. %

The state of a quantum system $A$ is given by a density operator $\rho$ on the Hilbert space $\Hset_A$.
Unless mentioned otherwise, we assume that the Hilbert spaces have finite dimensions.
A density operator is an Hermitian, positive semidefinite operator, with unit trace, \ie 
 $\rho^\dagger=\rho$, $\rho\succeq 0$, and $\trace(\rho)=1$.
The state is said to be pure if $\rho=\kb{\psi}$, for some vector $|\psi\rangle\in\Hset_A$, where
$\langle \psi |=(|\psi\rangle)^\dagger$. %
A measurement of a quantum system is any set of operators $\{\Lambda_j \}$ that forms a positive operator-valued measure (POVM), \ie
the operators are positive semi-definite and %
$\sum_j \Lambda_j=\identity$, where $\identity$ is the identity operator. %
According to the Born rule, if the system is in state $\rho$, then the probability of the measurement outcome $j$ is given by $p_A(j)=\trace(\Lambda_j \rho)$. The qubit Pauli basis is denoted by $\{\identity,\mathsf{X},\mathsf{Y},\mathsf{Z}\}$.

Define the quantum entropy of the density operator $\rho$ as
\begin{align}
H(\rho) \triangleq& -\trace[ \rho\log(\rho) ]
\end{align}
which is the same as the Shannon entropy %
associated with the eigenvalues of $\rho$.
Given a bipartite state $\sigma_{AB}$ on $\Hset_A\otimes \Hset_B$,  define the quantum mutual information by
\begin{align}
I(A;B)_\sigma=H(\sigma_A)+H(\sigma_B)-H(\sigma_{AB}) \,. %
\end{align} 
The conditional quantum entropy and mutual information are defined by
$H(A|B)_{\sigma}=H(\sigma_{AB})-H(\sigma_B)$ and
$I(A;B|C)_{\sigma}=H(A|C)_\sigma+H(B|C)_\sigma-H(A,B|C)_\sigma$, respectively.

A pure bipartite state %
is called \emph{entangled} if it cannot be expressed as the tensor product %
of two states %
in $\Hset_A$ and $\Hset_B$. %
The maximally entangled state %
between two systems %
of dimension $D$ %
is defined by
$%
| \Phi_{AB} \rangle = \frac{1}{\sqrt{D}} \sum_{j=0}^{D-1} |j\rangle_A\otimes |j\rangle_B %
$, where $\{ |j\rangle_A \}_{j=0}^{D-1}$ and $\{ |j\rangle_B \}_{j=0}^{D-1}$  %
are respective orthonormal bases. %
Note that $I(A;B)_{\kb{\Phi}}=2\cdot \log(D)$.

\subsection{Quantum Channels with Random Parameters}
\label{subsec:Qchannel}
A quantum channel maps a quantum state at the sender system to a quantum state at the receiver system. 
Here, we consider a model of channel uncertainty, where the channel is governed by a random parameter that \emph{changes over time}. %
Formally, let $\{ \channel_{A\rightarrow B}^{(s)} ,\, s\in\Sset \}$ be a collection of linear, completely positive, and trace-preserving (CPTP) maps, indexed by $s$, each corresponds to a quantum physical evolution. It is assumed that the channel has a product form, \ie 
if the systems $A^n=(A_1,\ldots,A_n)$ are sent through $n$ channel uses, then the input state $\rho_{A^n}$ undergoes the tensor product mapping
\begin{align}
\channel^{(s^n)}_{A^n\rightarrow B^n}\equiv \bigotimes_{i=1}^n \channel_{A\rightarrow B}^{(s_i)} \,.
\end{align}

We consider a quantum channel %
with a memoryless random parameter sequence, where the parameter sequence $(S_1,S_2,\ldots)$ is i.i.d. $\sim q(s)$. That is, the joint distribution of the parameter sequence is given by $\Pr(S^n=s^n)=q^n(s^n)\equiv\prod_{i=1}^n q(s_i)$. Therefore, without CSI, the input-output relation is
\begin{align}
\rho_{B^n}= \sum_{s^n\in\Sset^n} q^n(s^n) \channel^{(s^n)}_{A^n\rightarrow B^n } (\rho_{A^n})
= \left(\sum_{s\in\Sset} q(s) \channel^{(s)}_{A\rightarrow B} \right)^{\otimes n} (\rho_{A^n}) \,.
\label{eq:QinOutn}
\end{align}
The sender and the receiver are often referred to as Alice and Bob. 

Equivalently, the random-parameter quantum channel can be defined by a CPTP map $\channel_{SA\rightarrow B}$ with a bi-partite input, such that the component $S$ is a classical system in a given fixed state,  \ie
\begin{align}
\rho_S=\sum_{s\in\Sset} q(s) \kb{ s }
\end{align} 
where $\{|s\rangle\}_{s\in\Sset}$ is an orthonormal basis of the Hilbert space $\Hset_S$. Given CSI at the encoder, \ie when Alice  has access to the parameter sequence, or a part of it, then the input system $A^n$ can be correlated with $S^n$ as well.

We will also consider the quantum-classical special case. 
\begin{definition}
A measurement channel (or, q-c channel) $\Mset_{A\rightarrow Y}$ has the following form,
\begin{align}
\Mset_{A\rightarrow Y}(\rho_{A})=  \sum_{y\in\Yset} \trace( \Lambda_{y} \rho_A) \kb{y}
\end{align}
for some POVM $\{\Lambda_y \}$ and orthonormal vectors $\{ |y\rangle \}$. A random-parameter channel is called a measurement channel when the collection of CPTP maps consists of q-c channels. We denote the random-parameter measurement channel by %
$\Mset_{SA\rightarrow Y}$ to distinguish it from the general channel $\channel_{SA\rightarrow B}$. %
\end{definition}
A more general class of channels is that of entanglement-breaking channels. The definition is given below.
\begin{definition}
\label{def:EB}
A quantum channel $\mathcal{E}_{A\rightarrow B}$ is called entanglement breaking if for every  input state $\rho_{A A'}$, where $A'$ is an arbitrary reference system, the channel output 
is separable,  \ie
\begin{align}
( \mathcal{E}_{A\rightarrow B} \otimes \identity)(\rho_{A A'})= \sum_{x\in\Xset} p_X(x)  \psi_{B}^x \otimes \psi^x_{A'} 
\end{align}
for some probability distribution $p_X(x)$ and pure states $\psi_{B}^x$ and $\psi_{A'}^x$.   We say that a random-parameter channel $\channel_{SA\rightarrow B}$ is entanglement-breaking if
each $\channel^{(s)}_{A\rightarrow B}$ is entanglement breaking, for $s\in\Sset$.
\end{definition}
Every entanglement-breaking channel $\mathcal{E}_{A\rightarrow B}$  can be represented as a serial concatenation of a measurement channel followed by a classical-quantum channel \cite[Corollary 4.6.1]{Wilde:17b}. That is, if $\Eset_{A\rightarrow B}$ is entanglement breaking, then there exists a pair of channels, $\Pset_{Y\rightarrow B}$ and $\Mset_{A\rightarrow Y}$, such that
\begin{align}
\Eset_{A\rightarrow B}= \Pset_{Y\rightarrow B} \circ \Mset_{A\rightarrow Y}
\end{align}
where $Y$ is classical.

\begin{figure}
\begin{center}
\includegraphics[scale=0.6,trim={2cm 9cm 1cm 9cm},clip]{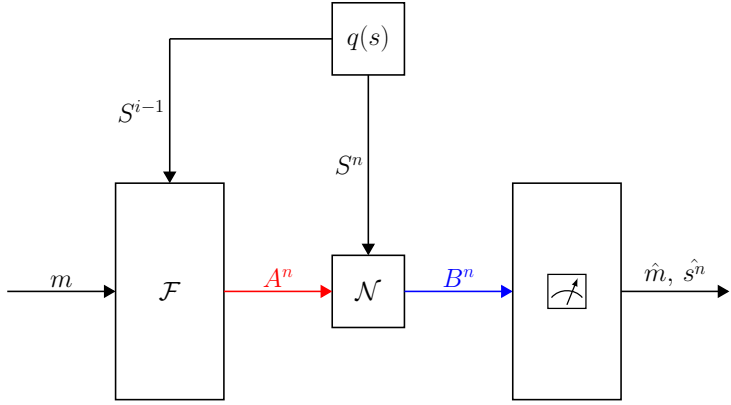} %
\end{center}
\caption{
Coding for a quantum channel $\channel_{SA\rightarrow B}$ that depends on a random parameter $S\sim q(s)$, with strictly-causal side information at the encoder and parameter estimation at the decoder. The quantum systems of Alice and Bob are marked in red and blue, respectively.
 Alice chooses a classical message $m$. At time $i$, given the parameter sequence $s^{i-1}$, %
her encoder $\Eset$ prepares a state $\rho^{m,s^{i-1}}_{ A_i}$, and then transmits the system $A_i$ over the quantum channel  $\channel_{SA\rightarrow B}$. 
 Bob receives the channel output systems $B^n$  and performs a measurement. The outcome is the estimated message $\hat{m}$ and reconstruction sequence $\hs^n$. 
With causal side information or non-causal side information, $S^{i-1}$ is replaced by $S^i$ or $S^n$, respectively.
}
\label{fig:siCode}
\end{figure}

\subsection{Coding}
\label{subsec:Mcoding}
We define %
a  code to transmit classical information. 
We will address four CSI scenarios.
With strictly-causal CSI, Alice knows, the \emph{past} random parameters $S^{i-1}$; given causal CSI, she knows the \emph{past and present} parameters $S^{i}$; with non-causal CSI, the entire sequence $S^n$ is available to her a priori; and without CSI, Alice is ignorant.
In all of those cases, Bob is unaware of the random parameters, and he has two tasks to perform. He is required to decode the message and to reconstruct the parameter sequence $S^n$ with a limited distortion.
Let %
$d:\Sset\times\widehat{\Sset}\rightarrow [0,\infty)$ be a bounded distortion function, with $d_{\max}\equiv \max_{s,\hs} d(s,\hs)$. Denote the average distortion between a parameter sequence $s^n$ and a reconstruction sequence $\hs^n$ by
\begin{align}
d^n(s^n,\hs^n)\triangleq \frac{1}{n} \sum_{i=1}^n d(s_i,\hs_i) \,.
\label{eq:dn}
\end{align}

\begin{definition} %
\label{def:EAcapacity}
A $(2^{nR},n)$  code with strictly-causal CSI at the encoder consists of the following:   
a message set $[1:2^{nR}]$, where $2^{nR}$ is assumed to be an integer, 
a sequence of encoding maps (channels) $\Fset^{(i)}_{M S^{i-1}\rightarrow A^i}$ for $i\in [1:n]$,
  and a decoding POVM $\{ \Lambda^{m,\hat{s}^n}_{B^n}  \}_{m\in [1:2^{nR}],\hat{s}^n\in\hat{\Sset}^n}$.
	The encoding maps must be consistent in the sense that the states $\rho_{A^i}^{m,s^{i-1}}\equiv 
	\Fset^{(i)}_{M,S^{i-1}\rightarrow A^i}
	(m,s^{i-1})$ satisfy $\trace_{A_{i+1}^n}(\rho_{A^n}^{m,s^{n-1}})=\rho_{A^i}^{m,s^{i-1}}$ for 
	$i\in [1:n]$.
We denote the code by $(\Fset,\Lambda)$.

The communication scheme is depicted in Figure~\ref{fig:siCode}.  
The sender Alice has the systems $A^n$ and the receiver Bob has the systems $B^n$. Alice chooses a classical message $m\in [1:2^{nR}]$. At time $i\in [1:n]$, Alice has the sequence of past parameters 
$s^{i-1}\in\Sset^{i-1}$, and can thus prepare the state 
$\rho^{m,s^{i-1}}_{A_i}$ and transmit the system $A_i$ over the channel %
$\channel_{SA\rightarrow B}$.

 Bob receives the channel output systems $B^n$ and performs the POVM
 $\{ \Lambda^{m,\hat{s}^n}_{B^n}  \}_{m\in [1:2^{nR}],\hat{s}^n\in\widehat{\Sset}^n}$. The conditional probability of decoding error, given that the message $m$ was sent, is given by 
\begin{align}
P_{e|m}^{(n)}(\Fset,\Lambda)= 
\trace\left[ \identity-\sum_{\hs^n\in\widehat{\Sset}^n} \Lambda^{m,\hat{s}^n}_{B^n} \sum_{s^n\in\Sset^n} q^n(s^n)
\channel^{(s^n)}_{A^n\rightarrow B^n}
(\rho_{A^n}^{m,s^{n-1}})%
 \right] \,.
\end{align}
The average distortion for the code $(\Fset,\Lambda)$ is 
\begin{align}
\Delta^{(n)}(\Fset,\Lambda)\triangleq  \sum_{s^n\in\Sset^n} \sum_{\hs^n\in\widehat{\Sset}^n}
d^n(s^n,\hs^n) \prob{S^n=s^n,\hat{S}^n= \hs^n}
\end{align}
where %
\begin{align}
\prob{S^n=s^n,\hat{S}^n= \hs^n}=
q^n(s^n) \cdot \frac{1}{2^{nR}} \sum_{m=1}^{2^{nR}} \sum_{\hat{m}=1}^{2^{nR}} \trace\Big[ \Lambda^{\hat{m},\hat{s}^n}_{B^n}
\channel^{(s^n)}_{A^n\rightarrow B^n}
(\rho_{A^n}^{m,s^{n-1}})%
\Big] \,.
\end{align}

A $(2^{nR},n,\eps,D)$ rate-distortion code satisfies 
$%
P_{e|m}^{(n)}(\Fset,\Lambda)\leq\eps $ %
for all $m\in [1:2^{nR}]$, 
and $\Delta^{(n)}(\Fset,\Lambda)\leq D$.  %
A rate $R>0$ is called achievable with distortion $D$  if for every $\eps>0$ and sufficiently large $n$, there exists a 
$(2^{nR},n,\eps,D)$
code. The capacity-distortion region $\opC_{\text{s-c}}(\channel)$ is defined as the set of achievable pairs $(R,D)$ with strictly-causal CSI. 

Alternatively, one may fix the average distortion constraint $D>0$ and consider the optimal transmission rate. The capacity-distortion function $C_{\text{s-c}}(\channel,D)$ is defined as the supremum of achievable rates $R$ for a given distortion $D$. 
Note that $C_{\text{s-c}}(\channel,d_{\max})$ reduces to the standard definition of the capacity of a quantum channel, without a distortion requirement or parameter estimation by the decoder.   
\end{definition}

We also address the causal and the non-causal setting. In the causal setting,  Alice has the present parameter value $S_i$ as well, and prepares $\rho^{m,s^n}_{A^n}$ such that $\rho^{m,s^n}_{A^i}\equiv \Eset^{(i)}_{MS^{i}\rightarrow A^i}(m,s^i)$.
Whereas, in the non-causal setting,  Alice has the entire parameter sequence $S^n$ a priori, and can thus prepare $\rho^{m,s^n}_{A^n}$ of any form. Without CSI, Alice sends a sequence in the state $\rho_{A^n}^{m}=\Eset_{M\rightarrow A^n}(m)$ that is  independent of the parameter sequence.
We use the subscripts `s-c', `caus', or `n-c' to indicate whether CSI is available at the encoder in a stictly-causal, causal, or non-causal manner, respectively.
The notation is summarized in the table in Figure~\ref{table:LcapacityNotation}.

\begin{figure*}[htb]	
\begin{center}
\hspace{-2cm}
\begin{tabular}{l|cccc}%
&$\;$ none $\;$ &	strictly-causal $\;$ &  causal 	 & non-causal   
								 \\	[0.2cm]   \hline \\			[-0.2cm]
Region					&	
										$\opC(\channel)$
									& $\opC_{\text{s-c}}(\channel)$ 	
									& $\opC_{\text{caus}}(\channel)$
									&	$\opC_{\text{n-c}}(\channel)$							
																																				\\[0.3cm] 
Function					&	
										$C(\channel,D)$
									& $C_{\text{s-c}}(\channel,D)$ 	
									& $C_{\text{caus}}(\channel,D)$
									&	$C_{\text{n-c}}(\channel,D)$										
\end{tabular}
\end{center}
  \caption{Notation of channel capacity-distortion regions and functions with and without CSI. The notation of the capacity-distortion regions is given in the first row, and of the capacity-distortion functions in the second row.
	The columns indicate the type of CSI that is available at the encoder. 
	}%
\label{table:LcapacityNotation}
\end{figure*}

\section{Related Work}
\label{subsec:Previous}
In this section, we briefly review known results for a quantum channel that does not depend on a random parameter and has no distortion constraint, \ie $\channel^{(s)}_{A\rightarrow B}=\Eset_{A\rightarrow B}$ for $s\in\Sset$, and $D=d_{\max}$. We also bring a general discussion on regularization, additivity, and entanglement-breaking channels.
We compare between Shor's original approach for single-letterization, based on additivity, and an alternative argument that follows from the methods by Wang \etal
\cite{WangDasWilde:17p}.
 In the sequel, we will use those observations in our capacity-distortion analysis in the absence of CSI.

\subsection{HSW Theorem}
\label{subsec:HSW}
Consider a channel $\Eset_{A\rightarrow B}$ without random parameters.
Define %
\begin{align}
\chi(\Eset)\triangleq   \max_{p_X(x), |\phi_A^x\rangle } I(X;B)_\rho  %
\label{eq:HolevoChan}
\end{align}
 with $\rho_{XB}\equiv \sum_{x\in\Xset} p_X(x) \kb{x}\otimes \Eset( \kb{ \phi_A^x })$ and 
$|\Xset|\leq |\Hset_A|^2$.
The objective functional $I(X;B)_\rho$ is referred to as the Holevo information with respect to the ensemble 
$\{ p_X(x), \Eset( \kb{ \phi_A^x })  \}$ and the channel $\Eset_{A\rightarrow B}$, while the formula $\chi(\Eset)$ itself is sometimes referred to as the Holevo information of the channel  \cite{Wilde:17b}.
 Next, we cite the HSW Theorem, which provides a regularized capacity formula for a quantum channel without parameters or distortion requirement.
\begin{theorem} [see {\cite{Holevo:98p,SchumacherWestmoreland:97p,Shor:02p}}]
\label{theo:CeaNoSI}
$\,$
\begin{enumerate}[1)]
\item
The capacity of a quantum channel $\Eset_{A\rightarrow B}$ without parameters is given by 
\begin{align}
C(\Eset,d_{\max})= \lim_{k\rightarrow \infty} \frac{1}{k} \chi \left( \Eset^{\otimes k} \right) \,.
\end{align}
\item
If $\Eset_{A\rightarrow B}$ is entanglement-breaking, then 
\begin{align}
C(\Eset,d_{\max})=  \chi  (\Eset) \,.
\end{align}
\end{enumerate}
\end{theorem}
In the second part of the lemma, we included Shor's result for the class of entanglement-breaking channels  \cite{Shor:02p} (see Definition~\ref{def:EB}). %
We note that this class includes both   classical-quantum channels and measurement channels. %
In particular, the capacity of a measurement channel
$\Mset_{A\rightarrow Y}^{(0)}$
without parameters is given by 
\begin{align}
C(\Mset^{(0)},d_{\max})= \max_{p_X(x), |\phi_A^x\rangle } I(X;Y) 
\end{align}
with $p_{Y|X}(y|x)=\langle \phi_A^x | \Lambda_y |\phi_A^x \rangle$.

\begin{remark}
\label{rem:noSI}
The setting of a random-parameter quantum channel $\channel_{S A\rightarrow B }$ \emph{without} side information and with $D=d_{\max}$ is equivalent to that of a channel that does not depend on a parameter, with %
$%
\Eset_{A\rightarrow B}=\sum_{s\in\Sset} q(s) \channel_{A\rightarrow B}^{(s)}
$ %
(see (\ref{eq:QinOutn})). On the other hand, with side information at the encoder, this equivalence does not hold, as the channel input is correlated with the parameter sequence.
\end{remark}

\subsection{Regularization} 
From a practical perspective, the Holevo information formula in (\ref{eq:HolevoChan}) is generally considered to be ``easy to compute",  given the channel statistics,  since there are efficient algorithms to solve this convex optimization problem numerically, as \eg in  \cite{RamakrishnanItenScholzBerta:20c}, up to a given precision and provided that the  dimensions of the Hilbert spaces, $\Hset_A$ and $\Hset_B$,  are not too large.
Yet, in Shannon theory, it is generally desirable to establish a single-letter computable capacity formula \cite{Korner:87b}. Beyond computability, the  disadvantage of a regularized multi-letter formula, of the form 
\begin{align}
\lim_{n\rightarrow\infty}\frac{1}{n}\mathsf{I}(\Eset^{\otimes n}) \,, %
\end{align}
 is that such characterization is not unique (see \cite[Section 13.1.3]{Wilde:17b}). Nonetheless, it should be emphasized that regularized characterizations are yet significant, since in many cases, the capacity can still be computed. Furthermore, there are interesting properties that can be derived even without a closed-form expression for the capacity \cite{PeregDeppeBoche:21p,SmithYard:08p}.

From a technical mathematical perspective, the difficulty in proving a single-letter converse part for a quantum channel is the hybrid nature of the Holevo mutual information $I(X;B)_\rho$, which involves a classical auxiliary variable $X$ and a quantum system $B$ (see Theorem~\ref{theo:CeaNoSI}).
Specifically, consider a channel $\Eset_{A\rightarrow B}$ without a random parameter.
From the familiar exercise of Fano's inequality and the chain rule,  one obtains the bound
\begin{align}
R-\eps_n &\leq %
\frac{1}{n} \sum_{i=1}^n  I(M;B_i|B^{i-1})_\rho 
\nonumber\\
&\leq \frac{1}{n} \sum_{i=1}^n  I(M,B^{i-1};B_i)_\rho 
\label{eq:ConvBoundH}
\end{align}
where $\eps_n$ tends to zero as $n\rightarrow\infty$ \cite{Holevo:98p,SchumacherWestmoreland:97p}.
In the attempt to establish an upper bound on the achievable rates in terms of the Holveo information $I(X;B)_\rho$, one is free to choose the auxiliary $X$  in the converse proof, in principle. Yet, $X$ needs to satisfy a certain Markov property, and more importantly in our discussion, $X$ must be classical.
Thereby, we cannot identify the auxiliary sequence $X_i$ with $(M,B^{i-1})$. We note that this stands in contrast to the entanglement-assisted capacity formula \cite{BennettShorSmolin:02p} \cite[Remark 5]{PeregDeppeBoche:21p}, where the auxiliary can be quantum.
A deeper perspective is given in Subsection~\ref{subsec:additivity} below.

\begin{remark}
One may look at the regularization problem from a different angle. 
In the book by Nielsen and Chuang \cite[Chapter 12]{NielsenChuang:02b}, 
the single-letter Holevo information is associated with the \emph{product-state capacity}.
Specifically, the authors consider a simplified setting where the encoder is constrained such that the channel input must be a product state. %
This means that not only entanglement is prohibited, but classical correlation is not allowed either. 
In this remark, we propose %
a more general encoding constraint, which makes more sense for a practical system.
For example, the model is suitable when the transmitter has access to multiple small or moderate-size quantum computers without interaction between them, where each computer has $b$ qubits.
In addition, in some qubit architectures, the physical limitations do not allow all qubits to ``talk to each other". 
That is, one cannot apply a quantum gate to any pair of qubits, but only to qubits that are at certain proximity to each other.
In order to account for those limitations, we impose the following encoding constraint.
Assume that the encoder's quantum systems $A^n$ are partitioned into sub-blocks of size $b$, such that the input state has the form
\begin{align}
\rho_{A^n}=  \rho_{A_1^b} \otimes \rho_{A_{b+1}^{2b}} \otimes \cdots  \otimes \rho_{A_{(\ell-1)b+1}^{n}}
\end{align}
with $\ell\equiv \frac{n}{b}$. 
As usual, the capacity $C_b(\Eset)$ under encoding constraint $b>0$ is defined as the supremum of the achievable rates with codes that satisfy the constraint above. %
Following the lines of \cite[Chapter 12]{NielsenChuang:02b}, it is can be shown that the 
capacity of a quantum channel $\Eset_{A\rightarrow B}$, without parameters, under encoding constraint $b>0$, is given by 
\begin{align}
C_b(\Eset)=\frac{1}{b}\chi(\Eset^{\otimes b})
\label{eq:Cb}
\end{align}
where
$%
\chi(\Eset)\equiv \max_{p_X, |\phi_A^x\rangle} I(X;B)_\rho
$ %
 is the Holevo information of the channel $\Eset_{A\rightarrow B}$.
Observe that this capacity formula  is computable, since $b>0$ is assumed to be a small constant. One may think of the formula on the RHS
of (\ref{eq:Cb}) as finite regularization.  By taking the limit $b\rightarrow \infty$, we recover the HSW theorem without encoding constraints.
\end{remark}

\subsection{Additivity}
\label{subsec:additivity}
Additivity is a central problem in the field of quantum Shannon theory \cite{Holevo:98p}. An information measure $\mathsf{I}(\Eset)$ is called additive
if the information of a product of two channels is equal to the sum of the respective informations. That is, for every pair of channels $\Eset$ and $\Gset$, 
\begin{align}
\mathsf{I}(\Eset\otimes\Gset)=\mathsf{I}(\Eset)+\mathsf{I}(\Gset) \,.
\label{eq:additivepair}
\end{align} 
It is well-known that this property holds for the Shannon capacity formula of a classical channel $\Eset_{X\rightarrow Y}$. 
The merit of this property is that regularized capacity formulas reduce to a single-letter computable formula when the corresponding information measure is additive.

For more than a decade, it was believed by many researchers that the Holevo information $\chi(\Eset)$, %
as defined in (\ref{eq:HolevoChan}), is also additive and that entanglement between input states does not increase the classical capacity of a quantum channel \cite[p. 554]{NielsenChuang:02b}. 
If the Holevo information of a channel is additive,
 then the regularization in the HSW characterization can be removed and the capacity can be expressed as $C(\Eset,d_{\max})=  \chi  (\Eset)$ (see Theorem~\ref{theo:CeaNoSI}). In fact, Fukuda and Wolf \cite{FukudaWolf:07p} established that $n$-fold additivity of the Holevo information is equivalent to its pairwise additivity. That is, when considering $\mathsf{I}(\Eset)=\chi  (\Eset)$, we have that
 $\mathsf{I}(\Eset^{\otimes n})=n\cdot \mathsf{I} (\Eset)$ holds for every quantum channel $\Eset_{A\rightarrow B}$ \emph{if and only if} (\ref{eq:additivepair}) holds for every pair of quantum channels $\Eset_{A_1\rightarrow B_1}$ and $\Gset_{A_2\rightarrow B_2}$.
Nevertheless, the additivity conjecture has  been refuted as Hastings \cite{Hastings:09p} demonstrated strict super-additivity of quantum channels in 2009.
That is, it was shown that there exist two channels $\Eset_{A_1\rightarrow B_1}$ and $\Gset_{A_2\rightarrow B_2}$  such that the Holevo informations satisfy
$\chi(\Eset\otimes\Gset)>\chi(\Eset)+\chi(\Gset)$.

\begin{remark}
The super-additivity of the Holevo information implies that the Holevo information does not provide a full characterization of the capacity, \ie $C(\Eset,d_{\max})\neq \chi(\Eset)$ in general. However,  it does \emph{not} imply that the operational capacity can be super-additive.  
As pointed out in  \cite[Section 8.4]{Holevo:12b}, it is an open problem whether there exist two channels $\Eset_{A_1\rightarrow B_1}$ and $\Gset_{A_2\rightarrow B_2}$  such that the operational capacity satisfies
$C(\Eset\otimes\Gset,d_{\max})>C(\Eset,d_{\max})+C(\Gset,d_{\max})$.
\end{remark}

\subsection{Entanglement-Breaking Channels}
\label{subsec:EBchannels}
Given the HSW characterization, it is straightforward to obtain a single-letter formula for measurement channels and classical-quantum channels. 
Shor \cite{Shor:02p} considered the more general class of entanglement-breaking channels, which includes both measurement and classical-quantum channels.
To obtain a single-letter characterization, Shor \cite{Shor:02p} has shown that the Holevo information of an entanglement-breaking channel is additive.
  On the other hand, we do not show additivity, but rather extend the methods of Wang \etal
\cite{WangDasWilde:17p}, and prove the converse part in a more direct manner.
We note that Shor's approach in \cite{Shor:02p} has more insight than ours, as it characterizes the fundamental properties of an entanglement-breaking channel. Yet, we believe that the alternative argument is easier to extend to more complex models, including channel uncertainty.

First, we demonstrate this argument for a channel $\Eset_{A\rightarrow B}$ without parameters. Consider the bound in 
(\ref{eq:ConvBoundH}).
As mentioned in Subsection~\ref{subsec:Qchannel}, if $\Eset_{A\rightarrow B}$ is an entanglement-breaking channel, then it can be presented as a concatenation of a measurement channel, followed by a state-preparation channel, i.e.
\begin{align}
\Eset_{A\rightarrow B}= \Pset_{Y\rightarrow B}\circ \Mset_{A\rightarrow Y}
\end{align}
where $Y$ is classical. Therefore, by the quantum data processing theorem due to Schumacher and Nielsen \cite{SchumacherNielsen:96p}\cite[Theorem 11.9.4]{Wilde:17b}, $I(M,B^{i-1};B_i)_\rho \leq I(M,Y^{i-1};B_i)_\rho $. Since the sequence $Y^{n-1}$ is classical, we can identify the auxiliary sequence as $X_i=(M,Y^{i-1})$, hence
\begin{align}
R-\eps_n \leq \frac{1}{n} \sum_{i=1}^n I(X_i;B_i)_\rho %
\end{align}
which is bounded by the single-letter Holevo information of the channel.

In the sequel, we will use this argument to establish a single-letter characterization of the capacity-distortion region in the absence of CSI %
(see Subsection~\ref{subsec:noCSI} and Part 2 of Appendix~\ref{app:mainNone}).

\section{Main Results} 
\label{sec:main}
We state our results on the random-parameter quantum channel $\channel_{S A\rightarrow B}$ with and without  CSI at the encoder.
The analysis is based on the information-theoretic tools that are presented in Appendix~\ref{sec:Itools}.

\subsection{Strictly-Causal Side Information}
\label{subsec:mainSC}
We begin with our main result on the random-parameter quantum channel with strictly-causal CSI. That is, at time $i$, Alice has access to the parameters of the \emph{past}, $S^{i-1}$.
Define the rate-distortion region %
\begin{align}
\label{eq:inCeaSCausal}
\mathcal{R}_{\text{s-c}}(\channel)\triangleq \bigcup %
\left\{
\begin{array}{lrl}
(R,D) \,:\; & R \leq& I(Z,X;B)_\rho - I(Z;S|X) \\
						& D \geq& \sum\limits_{s,\hs,x,z} q(s)p_X(x) p_{Z|X,S}(z|x,s) %
						\trace( \Gamma_{B|x,z}^{\hs} \rho^{s,x}_{B} ) d(s,\hs)
\end{array}
\right\}
\end{align}
where the union is over the set of all distributions $p_X(x)p_{Z|X,S}(z|x,s)$, state collection $\{ \theta^{x}_{A}  \}$, and set of POVMs 
$\{ \Gamma_{B|x,z}^{\hs} \}$,
with
\begin{align}
\label{eq:StateMaxSC}
&\rho^{s,x}_{B} =  \channel^{(s)}_{A\rightarrow B}(\theta^{x}_{A}) \\
&\rho_{SZXB}=  \sum_{s\in\Sset}  \sum_{z\in\Zset}  \sum_{x\in\Xset} q(s) p_X(x)p_{Z|X,S}(z|x,s)  \kb{s} \otimes \kb{z} \otimes \kb{x} \otimes \rho^{s,x}_{B} %
 \,. \label{eq:ICscDist0}
\end{align}
Before we state the capacity-distortion theorem, we give the following lemma. In principle, one may use the property below in order to compute the region $\mathcal{R}_{\text{s-c}}(\channel)$ for a given channel.
\begin{lemma}
\label{lemm:pureCeaSCausal}
The union in (\ref{eq:inCeaSCausal}) can be restricted to pure states $ \theta^{x}_{A}= \kb{ \phi^{x}_{A} } $,
with $|\Xset|\leq |\Hset_A|^2+1$ and $|\Zset|\leq|\Hset_A|^2+|\Sset|$.
\end{lemma}
The restriction to pure states %
follows by state purification, and %
the cardinality bounds are based on the Fenchel-Eggleston-Carath\'eodory lemma \cite{Eggleston:66p}, using similar arguments as 
in \cite{YardHaydenDevetak:08p}.
 The details are given in 
Appendix~\ref{app:pureCeaSCausal}.

Our main result is given below.
\begin{theorem}
\label{theo:mainSC}
$\,$
\begin{enumerate}[1)]
\item
The capacity-distortion region of a random-parameter quantum channel $\channel_{S A\rightarrow B}$  with strictly-causal CSI at the encoder is given by 
\begin{align}
\opC_{\text{s-c}}(\channel)=  \bigcup_{k=1}^{\infty} \frac{1}{k} \mathcal{R}_{\text{s-c}}(\channel^{\otimes k})
 \,.
\end{align}
\item
For a random-parameter measurement channel $\Mset_{S A\rightarrow Y}$,
\begin{align}
\opC_{\text{s-c}}(\Mset)=  \mathcal{R}_{\text{s-c}}(\Mset) \,.
\end{align}
\end{enumerate}
\end{theorem}
The proof of Theorem~\ref{theo:mainSC} is given in Appendix~\ref{app:mainSC}.
To prove achievability, we extend  the classical block Markov coding to the quantum setting, and then apply the quantum packing lemma for decoding the message, and the classical covering lemma for the reconstruction of the parameter sequence. The gentle measurement lemma \cite{Winter:99p} %
alleviates the proof, as it guarantees that multiple decoding measurements can be performed without ``destroying" the quantum state, \ie such that the output state after each measurement is almost the same.

\begin{remark}
\label{rem:boundHolevoSC}
Observe that the bound on the rate in the definition of $\mathcal{R}_{\text{s-c}}(\channel) $ in (\ref{eq:inCeaSCausal})
can also be written as
\begin{align}
R\leq I(X;B)_\rho-\left[ I(Z;S|X)-I(Z;B|X)_\rho \right] \,,
\label{eq:rateBsc}
\end{align}
by the mutual information chain rule. As the expression in the square brackets above is nonnegative, by the data processing inequality, it follows that the rate is bounded by the capacity of the channel without CSI, \ie the Holevo information (see Subsection~\ref{subsec:HSW}). We will come back to this point  when we consider the capacity when the receiver is not required to estimate the channel parameters in Subsection~\ref{subsec:SInoEst}.
\end{remark}

\begin{remark}
The expression in the square brackets in (\ref{eq:rateBsc}) can be interpreted as the penalty that the encoder pays for the transmission to contain a (partial) representation of the parameter sequence.
 Luo and Devetak \cite{LuoDevetak:09p} have considered the source compression setting with SSI. In their setting, a classical memoryless source $S^n$ is compressed, and then reconstructed at the decoder with distortion $D$, using quantum side information systems $B^n$. Based on their results, the rate-distortion function $r(D,\rho_{SB})$ for this source compression setting is given by  $r(D,\rho_{SB})=\lim_{k\rightarrow\infty} \frac{1}{k} \mathsf{r}(D,\rho_{SB}^{\otimes k})$, where
\begin{align}
\mathsf{r}(D,\rho_{SB})=\min_{\substack{ p_{\tilde{Z}|S}(z|s)\,,\; \{\Gamma_{B|z}^{\hs}\} \,:\;\\
 \sum\limits_{s,\hs,z} q(s) p_{\tilde{Z}|S}(z|s) %
						\trace( \Gamma_{B|z'}^{\hs} \rho^{s}_{B} ) d(s,\hs)\leq D}} 
						\left[ I(Z';S)-I(Z';B)_\rho \right]
\end{align} 
(see Theorem 4.2 in \cite{LuoDevetak:09p}). Thereby, in our setting, the encoder's penalty can be interpreted as the average compression rate of the parameter sequence with the channel output as the decoders’s SSI. The more $Z$ is correlated with the channel parameter $S$, the parameter estimation will be better, \ie with a lower distortion. Yet, the penalty may be larger, resulting in a lower communication rate. 
\end{remark}

We illustrate our results with the following example. We will use the dephasing channel as a running example, and come back to it in the next sections as well.

\begin{example}
\label{example:dephasingSC}
Consider a random-parameter dephasing channel that is specified by
\begin{align}
\channel^{(0)}(\rho)&=\rho\\
\channel^{(1)}(\rho)&=\mathsf{Z}\rho\mathsf{Z}
\end{align}
with  a binary random parameter $S\sim\text{Bernoulli}(\eps)$,  where $\eps\in [0,1]$ is a given constant, \ie $q(1)=1-q(0)=\eps$.
 In other words, given a parameter sequence $S^n$, the parameter $S_i$ acts as a switch that controls the phase flip operation at time $i$.
Observe that without CSI, the decoder receives the average output of the standard dephasing channel, \ie $\overline{\channel}(\rho)\equiv (1-\eps)\rho+\eps \mathsf{Z}\rho\mathsf{Z}$. Given CSI at the encoder, however, the input state is correlated with the channel parameters.
Furthermore, our decoder needs to recover the message and estimate whether there was a phase flip at each time.
The natural measure for the distortion  between the binary parameter sequence and its reconstruction is the following: $d(s,\hs)=s+\hs \mod 2$. Namely, $d(s,\hs)=1$ if $\hs\neq s$, and $d(s,\hs)=0$ if $\hs= s$.

Clearly, if the encoder sends a constant transmission  $|+\rangle\otimes \cdots\otimes |+\rangle$, then the decoder can determine whether there was a phase flip at each instance and recover the parameters without distortion. Yet, the rate is zero as well. 
On the other hand, by restricting the transmission to the computational basis, we can communicate without error, since the states $|0\rangle$ and $|1\rangle$ are unaffected by the phase flips.
Thereby, if one is not interested in parameter estimation, the rate $R=1$ can achieved.
Based on Theorem~\ref{theo:mainSC}, we show that the following rate-distortion region is achievable for the random-parameter dephasing channel with strictly-causal CSI at the encoder, %
\begin{align}
\opC_{\text{s-c}}(\channel)\supseteq
\bigcup_{0\leq\alpha\leq \frac{1}{2}}\left\{
\begin{array}{lrl}
(R,D) \,:\; & R \leq& 1-[h(\alpha*\eps)-h(\alpha)] \\
						& D \geq& \alpha
\end{array}
\right\} \,,
\end{align}
where $a*b=(1-a)b+a(1-b)$ denotes the binary convolution operation, and 
$h(x)=-x\log(x)-(1-x)\log(1-x)$ is the binary entropy function. 
Here, we see the tradeoff between the communication rate and the distortion. Taking $\alpha=\frac{1}{2}$, we obtain the maximal rate $R=1$, but the distortion $D=\frac{1}{2}$ is that of guessing by a coin flip. 
On the other hand, for $\alpha=0$, the channel parameters  are recovered without distortion, while the communication rate is bounded by $R= 1-h(\eps)$.
To obtain the achievable region above from Theorem~\ref{theo:mainSC}, consider $k=1$. Set  the distribution of the input ensemble as $X\sim\text{Bernoulli}\left(\frac{1}{2}\right)$  over the state collection $\{|0\rangle \,,\; |1\rangle \}$. Define $Z=X+S+V \mod 2$ with $V\sim\text{Bernoulli}\left(\alpha\right)$, such that $V$, $X$, and $S$ are statistically independent. %
Given $X$ and $Z$, %
the decoder estimates the channel parameter by $\hS=X+Z\mod 2=S+V\mod 2$.
This yields 
$I(X,Z;B)_\rho=H(B)_\rho-H(B|X,Z)_\rho=1-0=1$, $I(Z;S|X)=H(Z|X)-H(V)=h(\alpha*\eps)-h(\alpha)$, and $\mathbb{E} d(S,\hS)=%
\Pr(V=1)=\alpha$. 
\end{example}

Equivalently, we can characterize the capacity-distortion function.
\begin{corollary}
\label{coro:CDfSC}
$\,$
\begin{enumerate}[1)]
\item
The capacity-distortion function of a random-parameter quantum channel $\channel_{S A\rightarrow B}$  with strictly-causal CSI at the encoder is given by 
\begin{align}
C_{\text{s-c}}(\channel,D)=  \lim_{k\rightarrow \infty} \frac{1}{k}  \max_{ \substack{ p_{X^k}(x^k)p_{Z^k|X^k,S^k}(z^k|x^k,s^k) \,,\; \\ 
|\phi^{x^k}_{A^k} \rangle \,,\; 
\{ \Gamma_{B^k|x^k,z^k}^{\hs^k} \} \,:\;
\E d^k(S^k,\hS^k)\leq D
 }} [I(Z^k,X^k;B^k)_\rho - I(Z^k;S^k|X^k)]
\label{eq:Csc0}
\end{align}
with 
\begin{multline}
\rho_{S^k Z^k X^k B^k}=  \sum_{s^k,x^k,z^k} q^k(s^k) p_{X^k}(x^k)p_{Z^k|X^k,S^k}(z^k|x^k,s^k)  \kb{s^k} \\ \otimes \kb{z^k} \otimes \kb{x^k} 
\otimes \channel^{(s^k)}_{A^k\rightarrow B^k}( \kb{ \phi^{x^k}_{A^k} }) \,.
\end{multline}
\item
For a random-parameter measurement channel $\Mset_{S A\rightarrow Y}$, %
\begin{align}
C_{\text{s-c}}(\Mset,D)=  \max_{ \substack{ p_X(x)p_{Z|X,S}(z|x,s) \,,\; \\ 
|\phi^{x}_{A} \rangle \,,\; 
\{ \Gamma_{Y|x,z}^{\hs} \} \,:\;
\E d(S,\hS)\leq D
 }} [I(Z,X;Y) - I(Z;S|X)]
\end{align}
with $p_{Y|X,S}(y|x,s)=  \langle \phi^{x}_{A}|  \Lambda_{y}^{(s)}  |\phi^{x}_{A}\rangle$.
\end{enumerate}
\end{corollary}
The corollary follows from Lemma~\ref{lemm:pureCeaSCausal} and Theorem~\ref{theo:mainSC}. 
For example, following the derivation in Example~\ref{example:dephasingSC}, the capacity-distortion function of the random-parameter dephasing channel %
is bounded from below by
\begin{align}
C_{\text{s-c}}(\channel,D)\geq 1-[h(D*\eps)-h(D)]
\end{align}
for $0\leq D\leq \frac{1}{2}$.

\subsection{Causal Side Information}
Next, we consider the random-parameter quantum channel with causal CSI, where Alice has access to the  \emph{past and present} random parameters, \ie $S^{i-1}$ and $S_i$.
Define the rate-distortion region %
\begin{align}
\label{eq:inCeaCausal}
\mathcal{R}_{\text{caus}}(\channel,D)\triangleq \bigcup %
\left\{
\begin{array}{lrl}
(R,D) \,:\; & R \leq& I(Z,X;B)_\rho - I(Z;S|X) \\
						& D \geq& \sum\limits_{s,\hs,x,z} q(s)p_X(x) p_{Z|X,S}(z|x,s) %
						\trace( \Gamma_{B|x,z}^{\hs} \rho^{s,x}_{B} ) d(s,\hs)
\end{array}
\right\}
\end{align}
where the union is over the set of all distributions $p_X(x)p_{Z|X,S}(z|x,s)$, states $\{ \theta^{x}_{G} \}$, quantum channels 
$\Fset^{(s)}_{G\rightarrow A}$, and set of POVMs 
$\{ \Gamma_{B|x,z}^{\hs} \}$,
with
\begin{align}
\label{eq:StateMaxC}
&\eta^{x,s}_A= \Fset_{G\rightarrow A}^{(s)}(\theta^{x}_{G})  \\
&\rho^{s,x}_{B} =  \channel^{(s)}_{A\rightarrow B}(\eta^{s,x}_A) \\
&\rho_{SZXB}=  \sum_{s\in\Sset} \sum_{z\in\Zset}  \sum_{x\in\Xset}  q(s) p_X(x)p_{Z|X,S}(z|x,s)  \kb{s} \otimes \kb{z} \otimes \kb{x} \otimes \rho^{s,x}_{B} %
 \,. \label{eq:ICcDist0}
\end{align}
The union in (\ref{eq:inCeaSCausal}) can also be restricted to pure states $ \theta^{z,x}_{G}= \kb{ \phi^{z,x}_{G} } $ 
based on the 
same arguments as in the proof of Lemma~\ref{lemm:pureCeaSCausal}.

Observe that the difference between the characterizations with strictly-causal and causal  CSI, %
in $\mathcal{R}_{\text{s-c}}(\channel)$ and $\mathcal{R}_{\text{caus}}(\channel)$,
 respectively, 
is that the channel input   is 
$\theta_A^x$ in the former, and $\Fset_{G\rightarrow A}^{(s)}(\theta^{x}_{G})$ in the latter (\cf (\ref{eq:StateMaxSC}) and (\ref{eq:StateMaxC})).
That is, the input state here depends on the random parameter through the auxiliary channel $\Fset^{(s)}_{G\rightarrow A}$ in (\ref{eq:StateMaxC}). %
Further interpretation and intuition for the role of this auxiliary channel will be given in Remark~\ref{rem:ShannonStr} and 
Examples \ref{example:dephasingC} and \ref{example:depolC}.
Now, we give our main result on the random-parameter quantum channel with causal CSI.
\begin{theorem}
\label{theo:mainC}
$\,$
\begin{enumerate}[1)]
\item
The capacity-distortion region of a random-parameter quantum channel $\channel_{S A\rightarrow B}$  with causal CSI at the encoder is given by 
\begin{align}
\opC_{\text{caus}}(\channel)=  \bigcup_{k=1}^{\infty} \frac{1}{k} \mathcal{R}_{\text{caus}}(\channel^{\otimes k})
 \,.
\end{align}
\item
For a random-parameter measurement channel $\Mset_{S A\rightarrow Y}$,
\begin{align}
\opC_{\text{caus}}(\Mset)=  \mathcal{R}_{\text{caus}}(\Mset) \,.
\end{align}
\end{enumerate}
\end{theorem}
To prove achievability, we apply the %
 coding techniques from the proof of Theorem~\ref{theo:mainSC} to the virtual channel $\mathcal{V}^{(s)}_{G\rightarrow B}$, defined by 
\begin{align}
\mathcal{V}^{(s)}_{G\rightarrow B}(\rho_G)= \channel^{(s)}_{A\rightarrow B}\left( \Fset^{(s)}_{G\rightarrow A}(\rho_G)  \right) \,.
\end{align}
The proof outline for Theorem~\ref{theo:mainC} is given in Appendix~\ref{app:mainC}.

\begin{remark}
\label{rem:ShannonStr}
Shannon \cite{Shannon:58p} has shown that the capacity of a random-parameter classical channel $\Wset_{SX\rightarrow Y}$ with causal CSI at the encoder is given by 
 \begin{align}
C_{\text{caus}}(\Wset,d_{\max})=\max_{p_{T}} I(T;Y)
\label{eq:CClcaual}
\end{align}
with $X=T(S)$, where  $T:\Sset\rightarrow \Xset$ is an auxiliary function, which is commonly referred to as a \emph{Shannon strategy}. The maximization in the formula above is over the distribution of the Shannon strategy. In Shannon's achievability scheme, the strategy maps a parameter value $S_i=s$ to a classical input $X_i=T(s)$ \cite[Remark 7.6]{ElGamalKim:11b}.
 The auxiliary random-parameter channel $\mathcal{F}^{(s)}$ in (\ref{eq:StateMaxC}) can be viewed as the quantum counterpart of the
classical Shannon-strategy  (see discussions in previous work by the author \cite[Section IV.D]{Pereg:19a} on further relations between Shannon strategies and %
quantum channels with causal CSI). The effect of the quantum Shannon strategy is demonstrated in the examples below.
\end{remark}

In a similar manner as in the previous subsection, we can equivalently  characterize the capacity-distortion function, $C_{\text{caus}}(\channel,D)$.
We omit this characterization to save space.
Causal CSI may lead to a significant advantage compared to strictly-causal CSI, as demonstrated in the example below.

\begin{example}
\label{example:dephasingC}
Consider the random-parameter qubit dephasing channel from Example~\ref{example:dephasingSC}. Observe that knowing the current parameter $S_i$, at time $i$, the encoder can revert the dephasing using the following strategy:  Perform
$\Fset^{(0)}(\rho)=\channel^{(0)}(\rho)=\rho$ and $\Fset^{(1)}(\rho)=\channel^{(1)}(\rho)=\mathsf{Z}\rho\mathsf{Z}$. 
By Theorem~\ref{theo:mainC}, the capacity-distortion region of the random-parameter dephasing channel with causal CSI is given by
\begin{align}
\opC_{\text{caus}}(\channel)=
\left\{
\begin{array}{lrl}
(R,D) \,:\; & R \leq& 1 \\
						& D \geq& 0
\end{array}
\right\} \,.
\end{align}
The converse part is immediate, since the classical transmission rate over a qubit channel is always bounded by $1$.
To show the direct part using Theorem~\ref{theo:mainC}, consider $k=1$. Set $X\sim\text{Bernoulli}\left(\frac{1}{2}\right)$  over the input ensemble $\{|\pm\rangle \}$, and $Z\equiv 0 $. The decoder performs a measurement in the $\pm$-basis. Given $X$ and a measurement outcome $Y$, choose $\hS=Y+X\mod 2=S$. 
\end{example}

\begin{example}
\label{example:depolC}
Consider a random-parameter qubit depolarizing channel that is specified by
\begin{align}
\channel^{(0)}(\rho)&=\rho\\
\channel^{(1)}(\rho)&=\mathsf{X}\rho\mathsf{X}\\
\channel^{(2)}(\rho)&=\mathsf{Y}\rho\mathsf{Y}\\
\channel^{(3)}(\rho)&=\mathsf{Z}\rho\mathsf{Z}
\end{align}
with the following parameter distribution, 
\begin{align}
q(0)=1-\eps \,,\; q(1)=q(2)=q(3)=\frac{\eps}{3}
\end{align}
 where $\eps\in [0,\frac{3}{8}]$ is a given constant. In other words, %
the parameter $S_i$ chooses a Pauli operator that is applied to the $i$th input system. 
We note that without CSI, the average channel is the same as the standard depolarizing channel, \ie
\begin{align}
\overline{\channel}_{A\rightarrow B}(\rho)&\equiv \sum_s q(s)\channel^{(s)}(\rho)
\nonumber\\
&=(1-\eps)\rho+\frac{\eps}{3}\left( \mathsf{X}\rho\mathsf{X}+\mathsf{Y}\rho\mathsf{Y}+\mathsf{Z}\rho\mathsf{Z}  \right)
\nonumber\\
&=(1-p)\rho+p\pi
\end{align}
 where $\pi=\frac{\identity}{2}$ is the maximally mixed state, and $p\equiv \frac{4\eps}{3}$ is interpreted as the probability of depolarization (see \cite[Section 4.7.4]{Wilde:17b}).
Here, the decoder needs to recover the message and estimate which Pauli operator was applied. %
For the distortion to be measured by the Hamming distance between the parameter sequence and its reconstruction, let $d(s,\hs)=1$ if $\hs\neq s$, and $d(s,\hs)=0$ if $\hs= s$.

Knowing the current parameter $S_i$, at time $i$, the encoder can revert the operation of the channel using the following strategy:  Perform
$\Fset^{(0)}(\rho)=\channel^{(0)}(\rho)=\rho$, $\Fset^{(1)}(\rho)=\channel^{(1)}(\rho)=\mathsf{X}\rho\mathsf{X}$, $\Fset^{(2)}(\rho)=\channel^{(2)}(\rho)=\mathsf{Y}\rho\mathsf{Y}$, and $\Fset^{(3)}(\rho)=\channel^{(3)}(\rho)=\mathsf{Z}\rho\mathsf{Z}$. %
Therefore, if one ignores the parameter estimation requirement, then the rate $R=1$ can be achieved.
By Theorem~\ref{theo:mainC}, the following region is achievable for the random-parameter depolarizing channel with causal CSI,
\begin{align}
\opC_{\text{caus}}(\channel)\supseteq 
\bigcup_{0\leq \alpha\leq \eps}
\left\{
\begin{array}{lrl}
(R,D) \,:\; & R \leq& 1-[H(1-\eps,\frac{\eps}{3},\frac{\eps}{3},\frac{\eps}{3})-H(1-\alpha,\frac{\alpha}{3},\frac{\alpha}{3},\frac{\alpha}{3})] \\
						& D \geq& \alpha
\end{array}
\right\} \,.
\end{align}
Once more, we see the tradeoff between the communication rate and the distortion.
If we want the transmission to describe the parameter sequence without distortion, then this costs $H(S)=H(1-\eps,\frac{\eps}{3},\frac{\eps}{3},\frac{\eps}{3})$. Thereby, taking $\alpha=0$, we achieve $R=1-H(1-\eps,\frac{\eps}{3},\frac{\eps}{3},\frac{\eps}{3})$ and $D=0$.
At the other extreme, taking $\alpha=\eps$, we obtain the maximal rate 
$R=1$, but the distortion $D=\eps$ is that of ignorantly guessing `$0,0,\ldots,0$'.

To show achievability of the region above by Theorem~\ref{theo:mainC}, set the distribution of the input ensemble as $X\sim\text{Bernoulli}\left(\frac{1}{2}\right)$  over a qubit basis, and apply $\Fset^{(s)}$ to the basis vectors as specified above.
Let $\tS\sim (1-\alpha,\frac{\alpha}{3},\frac{\alpha}{3},\frac{\alpha}{3})$ and $T%
$ be statistically independent random variables such that 
 $S=\tS+T \mod4$, for some $0\leq \alpha%
\leq \eps$.
Define $Z=X+S+\tS \mod 4$.
Given $X$ and $Z$, the decoder chooses $\hS=Z-X\mod 4=S+\tS\mod 4$.
This yields 
$I(X,Z;B)_\rho=H(B)_\rho-H(B|X,S+\tS)_\rho=1-0=1$, $I(Z;S|X)=H(S)-H(S|T)=H(S)-H(\tS)=H(1-\eps,\frac{\eps}{3},\frac{\eps}{3},\frac{\eps}{3})-H(1-\alpha,\frac{\alpha}{3},\frac{\alpha}{3},\frac{\alpha}{3})$, and $\mathbb{E} d(S,\hS)=1-\Pr(Z-X-S\mod 4=0)=1-\Pr(\tS=0)=\alpha$. 
\end{example}

\subsection{Non-Causal Side Information}
We consider the random-parameter quantum channel with non-causal CSI.
Define the rate-distortion region %
\begin{align}
\label{eq:inCeaCnc}
\mathcal{R}_{\text{n-c}}(\channel)\triangleq \bigcup %
\left\{
\begin{array}{lrl}
(R,D) \,:\; & R \leq& I(X;B)_\rho - I(X;S) \\
						& D \geq& \sum\limits_{s,\hs,x} q(s) p_{X|S}(x|s) %
						\trace( \Gamma_{B|x}^{\hs} \rho^{s,x}_{B} ) d(s,\hs)
\end{array}
\right\}
\end{align}
where the union is over the set of all distributions $p_{X|S}(x|s)$, states $\{ \theta^{x}_{A} \}$, %
and set of POVMs 
$\{ \Gamma_{B|x}^{\hs} \}$,
with
\begin{align}
\label{eq:StateMaxNC}
&\rho^{s,x}_{B} =  \channel^{(s)}_{A\rightarrow B}(\theta^{x}_A) \\
&\rho_{SXB}=  \sum_{s\in\Sset}  \sum_{x\in\Xset} q(s) p_{X|S}(x|s)  \kb{s}  \otimes \kb{x} \otimes \rho^{s,x}_{B} %
 \,. \label{eq:ICncDist0}
\end{align}
We note that here, as opposed to the previous characterizations, the auxiliary variable $X$ is allowed to depend on the random parameter $S$ (\cf
(\ref{eq:ICscDist0}), (\ref{eq:ICcDist0}), and (\ref{eq:ICncDist0})).
Our main result on the random-parameter quantum channel with non-causal CSI is given below.
\begin{theorem}
\label{theo:mainNC}
The capacity-distortion region of the random-parameter quantum channel $\channel_{S A\rightarrow B}$ with non-causal CSI at the encoder is given by 
\begin{align}
\opC_{\text{n-c}}(\channel)=  \bigcup_{k=1}^{\infty} \frac{1}{k} \mathcal{R}_{\text{n-c}}(\channel^{\otimes k}) \,.
\end{align}
\end{theorem}
The proof of Theorem~\ref{theo:mainNC} is given in Appendix~\ref{app:mainNC}.
 To prove achievability, we use an extension of  the classical binning technique \cite{GelfandPinsker:80p} to the quantum setting, and then apply the quantum packing lemma and the classical covering lemma.
We note that even for a classical channel, a single-letter  characterization of the optimal region with 
non-causal CSI is an open problem \cite{BrossLapidoth:18p}.
As in Subsection~\ref{subsec:mainSC}, we can write an equivalent characterization in terms of the capacity-distortion function, $C_{\text{n-c}}(\channel,D)$. 
We omit this to save space. We illustrate the results with a simple example below. 
 Theorem~\ref{theo:mainNC}  is also be the basis for the bosonic dirty-paper analysis in  Section~\ref{sec:bosonic}.

\begin{example}
\label{example:erNC}
Consider the following random-parameter qubit channel,  
\begin{align}
\channel^{(0)}(\rho)&=\rho\\
\channel^{(1)}(\rho)&=|\psi\rangle\langle\psi|
\end{align}
with $S\sim\text{Bernoulli}(\eps)$, a Hamming distortion function as in the previous examples, and a given state $|\psi\rangle$, in the same qubit space. 
 In other words, %
the parameter $S_i$ chooses whether the $i$th input system is projected onto $|\psi\rangle$. 
Ignoring the CSI at the encoder, the model resembles the quantum erasure channel \cite{BennettDiVincenzoSmolin:97p} (see also \cite[Section 20.4.3]{Wilde:17b}), except that the ``erasure state" is orthogonal to the qubit space, while $|\psi\rangle$ in the present example is in the same qubit space. 
Nonetheless, we note that if the decoder knows the locations where the state is projected, then this model is equivalent to the quantum erasure channel.
Without this knowledge at the decoder, it is less obvious.

By Theorem~\ref{theo:mainNC}, the following region is achievable for the random-parameter channel above with non-causal CSI,
\begin{align}
\opC_{\text{n-c}}(\channel)\supseteq 
\bigcup_{0\leq \alpha\leq \frac{1}{2}}
\left\{
\begin{array}{lrl}
(R,D) \,:\; & R \leq& (1-\eps)h(\alpha) \\
						& D \geq& (1-\eps)\alpha
\end{array}
\right\} \,.
\end{align}
Again, we can see the tradeoff between the communication rate and the distortion. Let $|\psi_\perp\rangle$ be an orthogonal state with respect to $|\psi\rangle$.
Clearly, if the encoder transmits  $|\psi\rangle$ when $S_i=1$, and $|\psi_\perp\rangle$ when $S_i=0$, then the decoder can  recover the parameters without distortion, by performing a measurement in the corresponding basis, \ie $\{|\psi\rangle, |\psi_\perp\rangle\}$. %
Indeed, for $\alpha=0$, we achieve $(R,D)=(0,0)$. 
On the other hand, taking $\alpha=\frac{1}{2}$, we obtain the maximal rate 
$R=1-\eps$, which is also the capacity of the quantum erasure channel.

To show this, note that the bound on the rate on the RHS of (\ref{eq:inCeaCnc}) can also be expressed as
\begin{align}
R\leq H(X|S)-H(X|B)_\rho \,.
\end{align}
Given non-causal CSI at the encoder, we can choose an auxiliary $X$ that depends on the channel parameter $S$.
Let the input ensemble be the basis $\{|\psi\rangle, |\psi_\perp\rangle\}$. %
 The input distribution is chosen as follows. Let $V\sim \text{Bernoulli}(\alpha)$ be statistically independent of $S$.
If $S=0$, set $X=V+1\mod 2$. Otherwise, if $S=1$, then $X=0$.
As for the decoder, given $X$, set $\hS=X+1 \mod 2$.
This yields 
$H(X|B)_\rho=0$ and $H(X|S)=(1-\eps)H(V)=
(1-\eps)h(\alpha)$, and 
$\mathbb{E} d(S,\hS)=\Pr(\hS\neq S)=\Pr(S=0,V=1)=%
(1-\eps)\alpha$. 
\end{example}

\subsection{In the Absence of Side Information}
\label{subsec:noCSI}
Consider the case where Alice does not have access to the parameter sequence, yet Bob is required to estimate the sequence with limited distortion. Given our previous analysis, the proof of a regularized formula in this case is straightforward. %
However, here we obtain a single letter formula not just for measurement channels, but for the whole class of entanglement-breaking channels. 
As opposed to Shor \cite{Shor:02p}, we do not show additivity (see Subsection~\ref{subsec:additivity}).
Instead, we prove the converse part in a more direct manner using the observations that we have presented in Subsection~\ref{subsec:EBchannels}, which extend the methods by Wang \etal
\cite{WangDasWilde:17p}.

We give our capacity-distortion theorem for the random-parameter quantum channel without CSI.
Define %
\begin{align}
\label{eq:inCeaCnone}
\mathcal{R}(\channel)\triangleq \bigcup %
\left\{
\begin{array}{lrl}
(R,D) \,:\; & R \leq& I(X;B)_\rho  \\
						& D \geq& \sum\limits_{s,\hs,x} q(s) p_{X}(x) %
						\trace( \Gamma_{B|x}^{\hs} \rho^{s,x}_{B} ) d(s,\hs)
\end{array}
\right\}
\end{align}
where the union is over the set of all distributions $p_{X}(x)$, states $\{ \theta^{x}_{A} \}$, and set of POVMs 
$\{ \Gamma_{B|x}^{\hs} \}$,
with
\begin{align}
\label{eq:StateMaxNone}
&\rho^{s,x}_{B} =  \channel^{(s)}_{A\rightarrow B}(\theta^{x}_A) \\
&\rho_{SXB}=  \sum_{s\in\Sset}  \sum_{x\in\Xset} q(s) p_{X}(x)  \kb{s}  \otimes \kb{x} \otimes \rho^{s,x}_{B} %
 \,. %
\end{align}
We note that the union in (\ref{eq:inCeaCnc}) can be restricted to pure states $ \theta^{x}_{A}= \kb{ \phi^{x}_{A} } $ with
$|\Xset|\leq |\Hset_A|^2+1$, based on the 
same arguments as in the proof of Lemma~\ref{lemm:pureCeaSCausal}. %
\begin{theorem}
\label{theo:mainNone}
$\,$
\begin{enumerate}[1)]
\item
The capacity-distortion region of a random-parameter quantum channel $\channel_{S A\rightarrow B}$  without CSI is given by 
\begin{align}
\opC(\channel)=  \bigcup_{k=1}^{\infty} \frac{1}{k} \mathcal{R}(\channel^{\otimes k})
 \,.
\end{align}
\item
If $\channel_{S A\rightarrow B}$ is entanglement-breaking, then
\begin{align}
\opC(\channel)=  \mathcal{R}(\channel) \,.
\end{align}
\end{enumerate}
\end{theorem}
The proof of Theorem~\ref{theo:mainNone} is given in Appendix~\ref{app:mainNone}.
The proof of the first part follows by similar arguments as for the previous results, while the proof of the second part
is based on our observations in Subsection~\ref{subsec:EBchannels}.
The characterization of the capacity-distortion function $C(\channel,D)$ follows as before.

We revisit our running examples of dephasing and depolarizing channels.

\begin{example}
\label{example:dephasing0}
Consider the random-parameter qubit dephasing channel from Examples~\ref{example:dephasingSC} and \ref{example:dephasingC}. 
By Theorem~\ref{theo:mainNone}, the capacity-distortion region of the random-parameter dephasing channel without CSI is given by
\begin{align}
\opC(\channel)=
\left\{
\begin{array}{lrl}
(R,D) \,:\; & R \leq& 1-h(\eps) \\
						& D \geq& 0
\end{array}
\right\} \,.
\end{align}
To show achievability, set the distribution of the input ensemble %
and measure the parameter estimate %
as before. Observe that the region above is the same as in  Example~\ref{example:dephasingSC}. That is, for this channel, the capacity-distortion region with strictly-causal CSI is the same as without CSI.
\end{example}

\begin{example}
\label{example:depol0}
Consider the random-parameter depolarizing channel from Example~\ref{example:depolC}.
The capacity-distortion region of the random-parameter depolarizing channel without CSI is bounded by
\begin{align}
\opC(\channel)\supseteq 
\left\{
\begin{array}{lrl}
(R,D) \,:\; & R \leq& 1-h(\frac{2\eps}{3}) \\
						& D \geq& \frac{2\eps}{3}
\end{array}
\right\} \,.
\end{align}
To derive this achievable region, set the distribution of the input ensemble as $X\sim\text{Bernoulli}\left(\frac{1}{2}\right)$  over the state collection $\{|+\rangle \,,\; |-\rangle \}$. The channel output is then the same as that of a dephasing channel, as in the previous example. As the dephasing corresponds to the Pauli operators $\mathsf{Y}$ and $\mathsf{Z}$, the ``dephasing probability" is $\eps_0\equiv \frac{2\eps}{3}$. The decoder performs a measurement in the $\pm$-basis. Denote the measurement outcome by $Y$. 
If $X\neq Y$, then the decoder knows that the channel operator was either $\mathsf{Y}$ or $\mathsf{Z}$.
Then, the decoder chooses $\hS$ to be either $2$ or $3$ with equal probability.
If $X=Y$, then the decoder knows that the channel operator is not a dephasing one, \ie either $\identity$ or $\mathsf{X}$, and 
 chooses $\hS=0$. Thus, the average distortion is
$\mathbb{E} d(S,\hS)=(1-\eps)\cdot 0+\frac{\eps}{3}\cdot 1+\frac{2\eps}{3}\cdot \frac{1}{2}$.
\end{example}

\subsection{Without Parameter Estimation}
\label{subsec:SInoEst}
We obtain the following results as direct consequences of 
Corollary~\ref{coro:CDfSC} and Theorems \ref{theo:mainC}-\ref{theo:mainNC}.
As mentioned, the standard definition of the capacity, \ie %
when parameter estimation is not required at the decoder, is equivalent to the capacity-distortion function for $D=d_{\max}$. 
Henceforth, we  use the term `the capacity of $\mathcal{N}_{S A\rightarrow B}$ without CSI' referring to $C(\mathcal{N},d_{max})$. Similarly, $C_{\text{s-c}}(\mathcal{N},d_{max})$, $C_{\text{caus}}(\mathcal{N},d_{max})$, and $C_{\text{n-c}}(\mathcal{N},d_{max})$ are the capacities with strictly-causal, causal, and non-causal CSI, respectively (see Figure~\ref{table:LcapacityNotation}).  

The next corollaries generalize the results of Boche \etal \cite{BocheCaiNotzel:16p} on classical-quantum channels with CSI.

\begin{corollary}
\label{coro:CDfSC2}
The capacity of a random-parameter quantum channel $\channel_{S A\rightarrow B}$  with strictly-causal CSI at the encoder is the same as without CSI, \ie $C_{\text{s-c}}(\channel,d_{max})=C(\channel,d_{max})=\chi(\channel)$.
\end{corollary}
The direct part is immediate, since the encoder can simply ignore the CSI, while the converse part follows from Remark~\ref{rem:boundHolevoSC} and the HSW capacity Theorem, Theorem~\ref{theo:CeaNoSI}, where CSI is not available.

Next, we consider causal CSI.
\begin{corollary}
\label{coro:CDfC2}
$\,$
\begin{enumerate}[1)]
\item
The capacity  of a random-parameter quantum channel $\channel_{S A\rightarrow B}$  with causal CSI at the encoder is given by 
\begin{align}
C_{\text{caus}}(\channel,d_{\max})=  \lim_{k\rightarrow \infty} \frac{1}{k}  \sup_{ \substack{ p_{X^k}(x^k) \,,\; \\ 
|\phi^{x^k}_{G^k} \rangle \,,\; \Fset^{(s^k)}_{G^k\rightarrow A^k} 
 }} I(X^k;B^k)_\rho 
\end{align}
with 
\begin{align}
\rho_{S^k  X^k B^k}=  \sum_{s^k,x^k} q^k(s^k) p_{X^k}(x^k)  \kb{s^k}   \otimes \kb{x^k} 
\otimes \channel^{(s^k)}_{A^k\rightarrow B^k}\left( \Fset^{(s^k)}_{G^k\rightarrow A^k}( \kb{ \phi^{x^k}_{G^k} }) \right) \,.
\end{align}
\item
For a random-parameter measurement channel $\Mset_{SA\rightarrow Y}$, %
\begin{align}
C_{\text{caus}}(\Mset,%
d_{\max})=  \sup_{  p_X(x)\,,\;  |\phi^{x}_{G} \rangle \,,\; \Fset^{(s)}_{G\rightarrow A} } I(X;Y) 
\end{align}
with $p_{Y|X,Z,S}(y|x,z,s)=  \trace\left(  \Lambda_y  \Fset^{(s)}(\kb{\phi^{x}_{G}}) \right)$.
\end{enumerate}
\end{corollary}
The direct part follows by taking $Z=\emptyset$,
and the converse part holds
by the argument that we made in Remark~\ref{rem:boundHolevoSC} for strictly-causal CSI. 
In particular, for the depolarizing channel in Example~\ref{example:depolC},  Corollary~\ref{coro:CDfSC2} and Corollary~\ref{coro:CDfC2} imply %
$C_{\text{s-c}}(\channel,d_{\max})=1-h\left(\frac{p}{2}\right)=1-h\left(\frac{2\eps}{3}\right)$ and $C_{\text{caus}}(\channel,d_{\max})=1$.
For the random-parameter quantum channel with non-causal CSI, we recover the result in \cite[Corollary 1, part (c)]{AnshuHayashiWarsi:20p}.
\begin{corollary}[{see also \cite{AnshuHayashiWarsi:20p}}]
\label{coro:CDfNC2}
The capacity of a random-parameter quantum channel $\channel_{S A\rightarrow B}$  with non-causal CSI at the encoder is given by 
\begin{align}
C_{\text{n-c}}(\channel,d_{\max})=  \lim_{k\rightarrow \infty} \frac{1}{k}  \sup_{  p_{X^k|S^k}(x^k|s^k) \,,\; 
\theta_{A^k}^{x^k,s^k}   } [I(X^k;B^k)_\rho - I(X^k;S^k)]
\end{align}
with 
\begin{align}
\rho_{S^k X^k B^k}=  \sum_{s^k,x^k} q^k(s^k) p_{X^k|S^k}(x^k|s^k)  \kb{s^k}   \otimes \kb{x^k} 
\otimes \channel^{(s^k)}_{A^k\rightarrow B^k}\left( \theta_{A^k}^{x^k,s^k}) \right) \,.
\end{align}
\end{corollary}
The statement above immediately follows from Theorem~\ref{theo:mainNC} as the distortion constraint is inactive for $D=d_{\max}$. 
We will use the last corollary in the bosonic dirty-paper analysis in  the next section. %

\section{Bosonic Dirty Paper Coding}
\label{sec:bosonic}
In this section, we address the special case of a single-mode bosonic channel with classical interference in the modulation and with non-causal side information at the transmitter, without parameter estimation at the receiver, \ie $D=d_{\max}$. %
In the analysis, we will use our result in Corollary~\ref{coro:CDfNC2}.

\subsection{Introduction}
Consider a \emph{classical} channel $W_{Y|X,S}$ with random parameters.
Given non-causal CSI at the encoder, the channel is known as the Gel'fand-Pinsker model \cite{GelfandPinsker:80p}. The capacity of this channel is given by \cite{HeegardElGamal:83p}
\begin{align}
C_{\text{n-c}}(W,d_{\max})=\max_{p_{U,X|S}} [I(U;Y)-I(U;S)]
\end{align}
where $U$ is an auxiliary random variable such that $U \Cbar (X,S) \Cbar Y$ form a Markov chain.  The characterization above can also be obtained from Corollary~\ref{coro:CDfNC2}.

A  random-parameter Gaussian channel is specified by the input-output relation $Y=X+Z+S$,
with a real-valued Gaussian noise $Z\sim 
\mathcal{N}_{\mathbb{R}}(0,\sigma_Z^2)$, an additive interference $S$ known to the transmitter, and  an input power constraint $P$.
A well-known result by Costa \cite{Costa:83p} is that the capacity of the random-parameter Gaussian channel is the same as if the interference is not there, i.e. $C(W)=\frac{1}{2}\log\left(1+\frac{P}{\sigma_Z^2} \right)$.
Given that $S^n$ is \emph{not} known to the receiver, it is far from obvious that the interference can be canceled out without sacrificing transmission power.
The trivial strategy is to send $X=U-S$, such that $U$ represents the transmitted information and is uncorrelated with $S$, resulting in an intereference-free output, $Y=U+Z$. However, in the Gaussian case, the power constraint must be accounted for. Thereby, the trivial strategy above wastes transmission power and can only achieve a rate of $R\leq \frac{1}{2}\log\left(1+\frac{\max(P-\sigma_S^2,0)}{\sigma_Z^2} \right) $, which is sub-optimal.

 The derivation of the capacity of the random-parameter Gaussian channel with non-causal CSI requires a more gentle approach, and it is based on Costa's dirty-paper coding strategy \cite{Costa:83p}: Set 
\begin{align}
U=X+t S
\label{eq:UtX}
\end{align} %
such that $X$ is statistically \emph{independent} of $S$. The optimal choice of the coefficient $t$ turns out to be the same as that of the %
minimum mean-square error (MMSE) estimator $\widehat{X}=t(X+Z)$ for $X$ given the noisy observation $(X+Z)$ (see \cite[Section 4.1]{KeshetSteinbergMerhav:07n}), namely, 
\begin{align}
t=\frac{P}{P+\sigma_Z^2} \,.
\label{eq:tMMSE}
\end{align}
Explicit code constructions based on lattice codes were proposed in \cite{PhilosofErezZamir:03c,EreztenBrink:05p} and references therein. Furthermore, efficient algorithms for practical implementation were presented,  based on state-of-the art polar codes \cite{SimaChen:16c,BeilinBrshtein:21p}, LDPC codes \cite{WangHe:09c,RegeBalachandranKangKK:16p,BochererLentnerCirinoSteiner:19a}, and so on.

The bosonic channel is a simple quantum-mechanical model for optical communication over free space or optical fibers 
\cite{WPGCRSL:12p,WildeHaydenGuha:12p}, and it can be viewed as the quantum counterpart of the classical channel with additive white Gaussian noise (AWGN).
 An optical communication system consists of a modulated source of photons, the optical channel, and an optical detector. %
For a single-mode bosonic channel, the channel input is an electromagnetic field mode with an annihilation operator $\ha$, and the output is another mode with the annihilation operator $\hb$. The 
input-output relation in the Heisenberg picture \cite{HolevoWerner:01p} is given by %
\begin{align}
\hb=\sqrt{\eta}\, \ha +\sqrt{1-\eta}\,\he
\end{align}
where $\he$ is associated with the environment noise and the parameter $\eta$  is %
the transmissivity, $0\leq \eta\leq 1$, which depends on the length of the optical fiber and its absorption length \cite{EisertWolf:05c} (see Figure~\ref{fig:BSp}). 
For a lossy bosonic channel, the noise mode $\he$ is in a Gibbs thermal state $\tau(N_E)$ which consists of a mixture of coherent states,  where
\begin{align}
\tau(N)\equiv \int_{\mathbb{C}} d^2 \alpha \frac{e^{-|\alpha|^2}}{\pi N} \kb{\alpha} %
\end{align}
given an average photon number $N\geq 0$. 
 Modulation is performed such that the unitary displacement operator $D(\alpha)=\exp(\alpha \ha^\dagger-\alpha^* \ha)$ is applied to the vacuum state $\kb{0}$ \cite{WPGCRSL:12p}. 

\subsection{Model and Results}
We consider the single-mode lossy bosonic channel with a coherent-state protocol and a non-ideal displacement operation in the modulation process:
\begin{align}
|\zeta \rangle =D(\alpha+s)|0\rangle %
=|\alpha+s\rangle \,.
\label{eq:Zetan0}
\end{align}
This model can viewed as the quantum counterpart of the classical random-parameter Gaussian channel.
Based on Costa's writing-on-dirty-paper result \cite{Costa:83p}, the effect of the channel parameter can be canceled even when the decoder has no side information, and regardless of the input power constraint.
For both homodyne and heterodyne detection with a coherent-state protocol, the model reduces to a classical channel with either real or complex-valued Gaussian noise. Thereby, by applying Costa's dirty-paper coding strategy, we observe that the effect of the classical interference can be canceled for those channels as well. Then, we consider the bosonic channel with joint detection, for which the classical results do not apply, and derive a dirty-paper coding lower bound. Furthermore, considering the special case of a pure-loss bosonic channel, we demonstrate that the optimal coefficient for dirty paper coding is not necessarily the MMSE estimator coefficient as in the classical setting. We denote the random-parameter bosonic channel by $\Bset$.
To simplify the notation, we use the short notation $C_{\text{n-c}}(\Bset)\equiv C_{\text{n-c}}(\Bset,d_{\max})$ for the capacity of the random-parameter bosonic channel, without parameter estimation.

We begin with homodyne and heterodyne detection.
A homodyne measurement of a quadrature observable is implemented in practice by combining the target quantum mode with an intense local oscillator at a 50:50 beam splitter, and measuring the  photocurrent difference of the outgoing modes using two photodetectors 
\cite{BraunsteinvanLoock:05p}. 
When homodyne detection is used with a coherent-state protocol, the resulting channel $\Bset_{\text{hom}}$ is the random-parameter classical Gaussian channel
\begin{align}
Y=\sqrt{\eta}(\alpha+S)+Z_{\text{hom}}
\end{align}
with  a real-valued Gaussian %
parameter $S\sim \mathcal{N}_{\mathbb{R}}(0,N_S)$ and %
noise $Z_{\text{hom}}\sim \mathcal{N}_{\mathbb{R}}\left(0,\frac{1}{4}\left[ 2(1-\eta)N_E+1 \right] \right)$ \cite{Guha:08z}. %
Using the dirty-paper coding scheme, we take $\alpha\sim \mathcal{N}_{\mathbb{R}}(0,N_A)$ and $U=\alpha+t_0 S$ with $t_0=\frac{N_A}{N_A+N_E}$, such that $\alpha$ and $S$ are uncorrelated.
The effect of the interference is thus removed, and the capacity is given by 
\begin{align}
C_{\text{n-c}}(\Bset_{\text{hom}})=\frac{1}{2}\log\left( 1+\frac{4\eta N_A}{2(1-\eta)N_E+1} \right) %
\end{align}
 as without interference. %

In heterodyne detection, two quadratures are measured by combining  the measured mode with a vacuum %
mode into a 50:50 beam splitter, and homodyning the quadratures of the outcome modes \cite{WPGCRSL:12p}.
Heterodyne detection is described by a random-parameter  channel $\Bset_\text{het}$ with complex-valued Gaussian noise, specified by
\begin{align}
Y=\sqrt{\eta}(\alpha+S)+Z_{\text{het}}
\end{align}
with complex-valued circularly-symmetric Gaussian random parameter $S\sim \mathcal{N}_{\mathbb{C}}(0,\frac{1}{2}N_S)$ and noise  $Z_{\text{het}}\sim \mathcal{N}_{\mathbb{C}}(0,\frac{1}{2}[(1-\eta)N_E+1])$ \cite{Guha:08z}. 
Similarly, we use dirty-paper coding with $\alpha\sim \mathcal{N}_{\mathbb{C}}(0,\frac{1}{2} N_A)$ and $U=\alpha+t_0 S$, %
achieving the capacity
\begin{align}
C_{\text{n-c}}(\Bset_{\text{het}})=\log\left( 1+\frac{\eta N_A}{(1-\eta)N_E+1} \right) %
\end{align}
 as without interference.

At last, we consider the case where the decoder can perform an arbitrary quantum measurement on the output systems $B_1,\ldots,B_n$ together.
For joint detection \cite{HolevoWerner:01p}, the channel does not have a classical description. 
Based on Corollary~\ref{coro:CDfNC2}, the capacity of a random-parameter quantum channel $\Bset$ with CSI at the transmitter,  is given by the regularized formula $C_{\text{n-c}}(\Bset)=\lim_{n\to\infty} \frac{1}{n}\inC_{\text{n-c}}(\Bset^{\otimes n})$, with 
\begin{align}
\inC_{\text{n-c}}(\Bset)=\sup_{p_{X|S}\,,\; \theta_A^{x,s} } [I(X;B)_\rho-I(X;S)] \,.
\end{align} 
Previously, we have assumed that the space dimensions are finite.
Yet, this result is now extended to the bosonic channel with infinite-dimension Hilbert spaces following the discretization limiting argument by Guha \etal \cite{GuhaShapiroErkmen:07p}. %
Using the dirty-paper coding strategy, %
we obtain the lower bound $C_{\text{n-c}}(\Bset_{\text{joint}})\geq \inR_{\text{DPC}}(t)$,  %
\begin{align}
\inR_{\text{DPC}}(t)&\equiv I(\gamma;B)-I(\gamma;S) \Big|_{\gamma=\alpha+tS}
\nonumber\\&
=g(\eta(N_A+N_S)+(1-\eta)N_E) %
- g\left(\frac{\eta (1-t)^2 N_A N_S}{N_A+t^2 N_S}+(1-\eta)N_E \right)
\nonumber\\& 
-\log\left( \frac{N_A+t^2 N_S}{N_A} \right)
\label{eq:DPClower}
\end{align}
where the subscript `DPC' stands for `dirty-paper coding', and 
$g(N)$ is the von Neumann entropy of the thermal state $\tau(N)$, %
\begin{align}
g(N)=\begin{cases}
(N+1)\log(N+1)-N\log(N) & N>0 \,,\\
0& N=0 \,.
\end{cases}
\end{align}
The second equality in (\ref{eq:DPClower}) holds since the channel input is associated with $\zeta\equiv \alpha+S=\gamma+(1-t)S$ (see (\ref{eq:Zetan0})), and the conditional variance of the channel parameter $S$ given $\gamma$ is
\begin{align*}
\text{var}(S|\gamma)=\left[1-\frac{ (\text{cov}(\gamma,S))^2 }{\text{var}(S)\text{var}(\gamma)} \right]\text{var}(S)
=\frac{N_A N_S}{N_A+t^2 N_S} \,.
\end{align*}

\vspace{-0.2cm}
\begin{center}
\begin{figure}[bt!]
\includegraphics[scale=0.5,trim={-5cm 0.35cm 0 0},clip]
{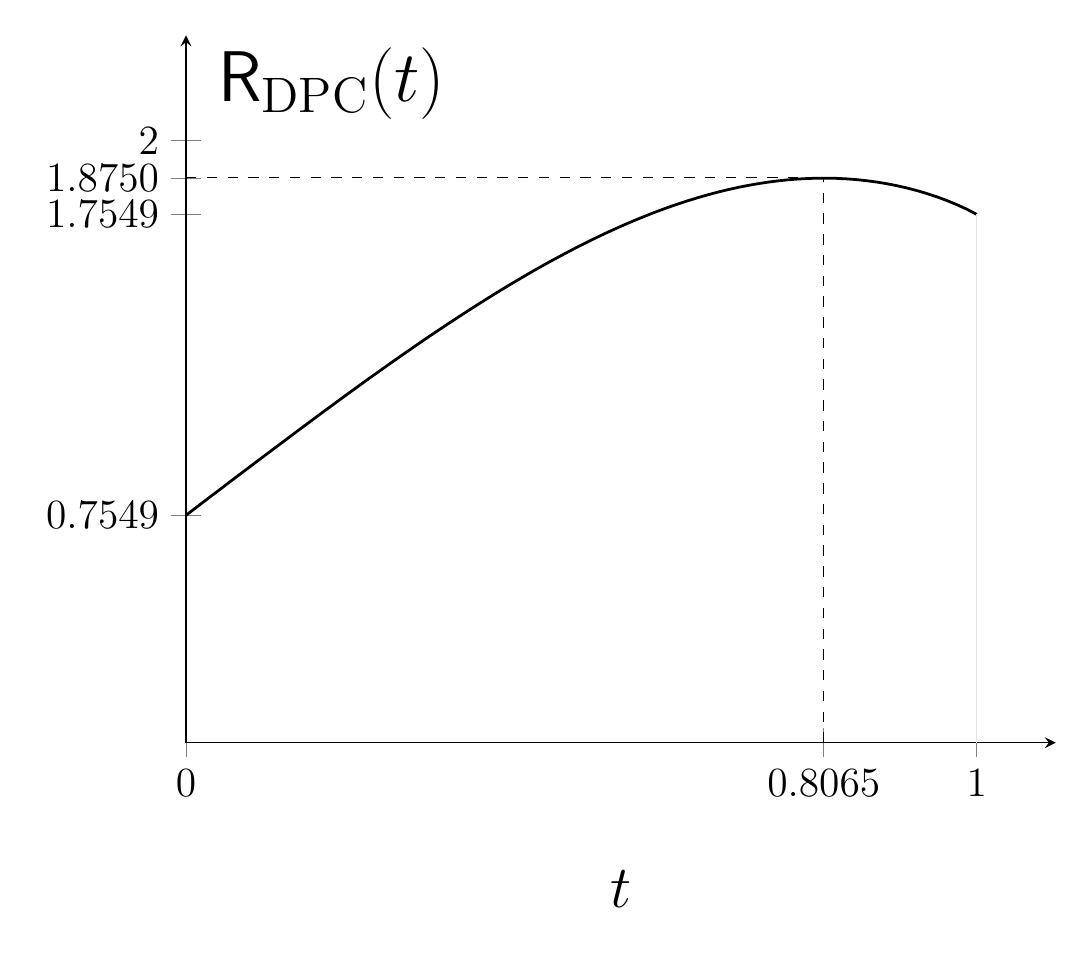} %
\caption{The dirty-paper coding lower bound for the pure-loss bosonic channel with joint detection and a coherent-state protocol.
}
\label{fig:jDPC}
\end{figure}
\end{center}

 In particular, consider the special case of a pure-loss bosonic channel, where $N_E= 0$. 
In this case, 
\begin{align}
\inR_{\text{DPC}}(t)=&
g(\eta(N_A+N_S))- g\left(\frac{\eta (1-t)^2 N_A N_S}{N_A+t^2 N_S} \right)%
-\log\left( \frac{N_A+t^2 N_S}{N_A} \right) \,.
\end{align}
To demonstrate, suppose that $N_A=N_S=2$ and $\eta=\frac{1}{2}$. Then, we have
\begin{align}
\inR_{\text{DPC}}(t)&=g(2)-g\left(\frac{(1-t)^2}{1+t^2}\right)-\log(1+t^2) \,. %
\end{align}
Ignoring the CSI, we obtain a rate $\inR_{\text{DPC}}(t=0)=g(2)-g(1)=%
3\log 3-4=0.7549$. 
Whereas, using the dirty-paper coding scheme with the MMSE coefficient $%
t_0=\frac{2}{2+0}=1$, we obtain a better rate:  $\inR_{\text{DPC}}(t=1)=3\log 3-2-\log 2=%
1.7549
$. The optimal value for dirty-paper coding turns out to be $t_{\max}=0.8065$, for which 
\begin{align}
\inR_{\text{DPC}}(t_{\max})=1.8750 .
\end{align}
 See Figure~\ref{fig:jDPC}. The rate above is higher than the homodyne and heterodyne-detection capacities, $C(\Eset_{\text{hom}})= 1.1609$ and $C(\Eset_{\text{het}})= 1$, respectively.
However, this rate is lower than 
 the joint-detection capacity without interference ($N_S=0$), which is given by 
$g(1)=  %
2$.

Our results can be further extended to other optical channels. In particular, the random-parameter thermal amplifier channel $\Aset$ with an amplification gain $\kappa>1$ has the input-output relation
$\hb=\sqrt{\kappa}\,\ha(s)+\sqrt{\kappa-1}\,\he^\dagger$ \cite{HolevoWerner:01p,GiovannettiGarciaCerfHolevo:14p}. In a similar manner, we obtain the dirty-paper coding lower bound %
\begin{align}%
C(\Aset)&\geq %
\max_{t\in [0,1]} \Bigg[ g(\kappa(N_A+N_S)+(\kappa-1)N_E) %
\nonumber\\&
- g\left(\frac{\kappa (1-t)^2 N_A N_S}{N_A+t^2 N_S}+(\kappa-1)N_E \right)
-\log\left( \frac{N_A+t^2 N_S}{N_A} \right) \Bigg].
\label{eq:DPClowerAmpl}
\end{align}%

\subsection{Concluding Remarks}
\label{subsec:remarks}
We conclude with the following remarks on the comparison between the classical and quantum dirty-paper settings:
\begin{enumerate}
\vspace{0.01cm}
\item
Costa \cite{Costa:83p} provided the intuitive analogy of `writing on dirty paper'. %
When a writer (Alice) is given a dirty paper, she knows the location and intensity of the dirt spots before writing. On the other hand, the reader (Bob) sees a mixture of the written text (channel input) and the dirt (channel parameter) without prior knowledge. %
In our setting, the dirt is the interference $s_i$ in the modulation displacement $D(\alpha_i+s_i)$. Alternatively, in the quantum watermarking scheme that we have described above, the dirt is the host covertext. 

\vspace{0.1cm}
\item
The classical capacity result can be derived using the dirty-paper coding strategy in (\ref{eq:UtX})-(\ref{eq:tMMSE}) following the observation that $U-tY=X-t(X+Z)$ is the error of the MMSE estimation of $X$ given $V=X+Z$, hence it is statistically independent of the observation. Thereby,  $(U-tY)$ is jointly independent of $(V,S)$. This, in turn, implies that $(U-tY)$ and $Y=V+S$ are statistically independent, leading to %
 $H(U|Y)=H(U-tY)=%
H(X|V)$ which can be used in order to show that
\begin{align} 
I(U;Y)-I(U;S)=H(U|S)-H(U|Y)=I(X;V) 
\end{align}
(see further details in \cite{KeshetSteinbergMerhav:07n} \cite[Section 7.7]{ElGamalKim:11b}). For a bosonic channel with joint detection, we can also write the capacity in terms of $%
H(X|S)-H(X|B)_\rho$, with conditioning on the channel output. However, conditioning on a quantum system does not necessarily carry the meaning of an observation as in the classical setting \cite{HorodeckiOppenheimWinter:05p}. %

\vspace{0.1cm}
\item
While dirty-paper coding was originally introduced to treat a channel with random parameters \cite{Costa:83p}, the technique is useful in multi-user setups of wireless communications as well, such as the multiple-input multiple-output (MIMO) broadcast fading channel \cite{WeingartenSteinbergShamai:06p}. It is only natural to apply and extend our results to multi-mode bosonic networks.

\end{enumerate}

\section{Summary and Discussion}
\label{sec:discussion}
We have considered a  quantum channel $\channel_{A\rightarrow B}^{(s)}$ that depends on a classical random parameter $S\sim q(s)$, when the decoder is required to reconstruct the parameter sequence in a lossy manner, \ie with limited distortion. This model can be viewed as the quantum analog of the classical rate-and-state (RnS) channel.

We consider two applications for this model: digital multicast using \emph{quantum} communication channels, and classical watermarking with a quantum embedding. 
The first application  is digital multicast, where the message represents digital control information that is multicast on top of an existing analog transmission, which is also estimated by the receiver.
 In the watermarking application, an authentication message $m$ is mixed within
 classical host data $S^n$ (``stegotext''), and this mixture is encoded into a quantum state that is sent to an authenticator.
The estimation of the channel parameters at the decoder corresponds to a scenario where
 the host data $S^n$ itself contains desirable information.
Our model can also be interpreted as a form of quantum metrology \cite{GiovannettiLloydMaccone:11p}, where the decoder performs  measurements on the received (quantum) systems in order to estimate classical noise parameters, while exploiting the entanglement generated by the encoder.

The scenarios that we studied in the present work include either strictly-causal, causal, or non-causal channel side information (CSI) available at the encoder, as well as the case where CSI is not available. 
With strictly-causal CSI, Alice knows, the \emph{past} random parameters $S^{i-1}$; given causal CSI, she knows the \emph{past and present} parameters $S^{i}$; with non-causal CSI, the entire sequence $S^n$ is available to her a priori; and without CSI, Alice is ignorant.
In all of those cases, Bob is unaware of the random parameters, and he has two tasks to perform. He is required to decode the message and to reconstruct the parameter sequence $S^n$ with a limited distortion, $D$.
We derived regularized formulas for the capacity-distortion tradeoff regions. %
In the special case of measurement channels,  single-letter characterizations were established for the strictly-causal and causal settings. %
Furthermore, in the more general case of entanglement-breaking channels, a single-letter characterization was derived when CSI is not available.
We also demonstrated the results in multiple examples, such as random-parameter dephasing channels and depolarizing channels.

While reviewing previous work in Section~\ref{subsec:Previous},
we reviewed single-letterization and regularization, additivity, and entanglement-breaking channels; and
we compared between Shor's original approach for single-letterization, based on additivity, and the alternative argument that follows from \cite{WangDasWilde:17p}. Later, we used this alternative argument in the analysis for our setting.
In particular, considering entanglement-breaking channels without CSI, we used a different approach from that of Shor \cite{Shor:02p}.
As opposed to Shor \cite{Shor:02p}, we did not show additivity of the capacity formula, but rather extended the methods of Wang \etal
\cite{WangDasWilde:17p} to prove the converse part in a more direct manner. 
This more direct approach has less insight compared to Shor's additivity argument, and yet, we believe that it can be easier to extend to complex settings,  as with parameter estimation at the decoder.

To prove achievability with strictly-causal CSI, we extended  the classical block Markov coding method from \cite{ChoudhuriKimMitra:13p} to the quantum setting, and then applied the quantum packing lemma for decoding the message, and the classical covering lemma for the reconstruction of the parameter sequence.
The gentle measurement lemma  alleviates the proof, as it guarantees that multiple decoding measurements can be performed without collapsing the quantum state and such that the output state after each measurement is almost the same. Thus, we can separate between measurements for recovering the message and for sequence reconstruction.
Achievability with causal CSI was proved using similar techniques with the addition of a quantum ``Shannon-strategy" encoding operation. %
To prove achievability with non-causal CSI, we used an extension of  the classical binning technique \cite{HeegardElGamal:83p} to the quantum setting. %

Furthermore, we introduced bosonic dirty-paper coding.  We considered the single-mode lossy bosonic channel with a coherent-state protocol and a non-ideal displacement operation in the modulation process.
The channel parameters in our model represent classical interference in the transmission equipment, which the transmitter becomes aware of, while the receiver is not. 
Alternatively, this can be viewed as a watermarking model with a quantum embedding.
Given a classical host data sequence $s_1,\ldots,s_n$, %
Alice encodes an authentication message $m$ into a watermark $(\alpha_i(m,s_1,\ldots,s_n))_{i=1}^n$. Next, Alice performs a quantum embedding of the watermark; 
she prepares a \emph{watermarked state} $|\zeta_1 \zeta_2 \cdots \zeta_n \rangle$ where $\zeta_i\equiv D(\alpha_i+s_i)|0\rangle=|\alpha_i+s_i\rangle$, %
and transmits it to the authenticator Bob through the optical fiber.
The capacity of the random-parameter bosonic channel represents the optimal rate at which the authenticator can recover the messages with high fidelity. %

First, we considered homodyne and heterodyne detection. Both of those settings reduce to a classical random-parameter channel with either real or complex-valued Gaussian noise. Thereby, we observed that based on Costa's dirty-paper solution, the effect of the classical interference can be canceled, and the capacity is the same regardless of the intensity of the interference. 
Then, we considered joint detection, in which case, the problem does not reduce to that of a classical description.
We derived a dirty-paper coding lower bound based on the results above, using an auxiliary $\gamma=\alpha+tS$ with a general coefficient $t\in [0,1]$, such that $\alpha\sim\mathcal{N}_{\mathbb{C}}(0,\frac{N_A}{2})$ is statistically independent of the channel parameter $S$.
Considering the special case of a pure-loss bosonic channel, we showed that the optimal coefficient is not necessarily the MMSE value $t_0=\frac{N_A}{N_A+N_E}$.

As a consequence of our main results, we obtained regularized formulas for the capacity of random-parameter  quantum channels with strictly-causal, causal, or non-causal CSI, %
generalizing the previous results by Boche \etal \cite{BocheCaiNotzel:16p} on classical-quantum channels.
We believe that this could open the door to extend other important classical side-information models to quantum communication.

\section{Acknowledgements}
The author gratefully thanks Roberto Ferrara (Technical University of Munich) for useful discussions.

This work was supported by the German BMBF Grant 16KIS0856 and the Israel CHE Fellowship for Quantum Science and Technology.

\begin{appendices}
\section{Information Theoretic Tools}
\label{sec:Itools}
To derive our results, we use the quantum version of the method of types properties and techniques. The basic definitions and lemmas that are used in this paper are given below.

\subsection{Classical Types}
The type of a classical sequence $x^n$ is defined as the empirical distribution $\hP_{x^n}(a)=N(a|x^n)/n$ for $a\in\Xset$, where $N(a|x^n)$ is the number of occurrences of the symbol $a$ in the sequence $x^n$. The set of all types over $\Xset$ is then denoted by 
$\pSpace_n(\Xset)$.
The type class associated with a type  $\hP\in \pSpace_n(\Xset)$ is defined as the set of sequences of that type, %
\ie
\begin{align} 
\Tset(\hP)\equiv\left\{ x^n\in\Xset^n \,:\; \hP_{x^n}=\hP  \right\} \,.
\end{align}
For a pair of sequences $x^n$ and $y^n$, we give similar definitions in terms of the joint type $\hP_{x^n,y^n}(a,b)=N(a,b|x^n,y^n)/n$ for $a\in\Xset$, $b\in\Yset$, where $N(a,b|x^n,y^n)$ is the number of occurrences of the symbol pair $(a,b)$ in the sequence 
$(x_i,y_i)_{i=1}^n$. Given a sequence $y^n\in \Yset^n$, we further define the conditional type $\hP_{x^n|y^n}(a|b)=N(a,b|x^n,y^n)/N(b|y^n)$
and the conditional type class  
\begin{align} 
\Tset(\hP|y^n)\equiv\left\{ x^n\in\Xset^n \,:\; \hP_{x^n,y^n}(a,b)=\hP_{y^n}(b)\hP(a|b) \right\} \,.
\end{align}

Given a probability distribution $p_X\in\pSpace(\Xset)$,  the $\delta$-typical set  is defined as
\begin{align}
\tset(p_X)\equiv \bigg\{ x^n\in\Xset^n \,:\; %
\left| \hP_{x^n}(a) - p_X(a) \right|\leq\delta \quad\text{if $\, p_X(a)>0$}&  \nonumber\\ 
 \hP_{x^n}(a)=0 \quad\text{if $\, p_X(a)=0$} &, \;\text{$\forall$ $a\in\Xset$} \bigg\}
\end{align}

The covering lemma is a powerful tool in classical information theory \cite{CsiszarKorner:82b}. 
\begin{lemma}[Classical Covering Lemma {\cite{CsiszarKorner:82b}\cite[Lemma 3.3]{ElGamalKim:11b}}]
\label{lemm:covering}
Let $X^n\sim \prod_{i=1}^n p_X(x_i)$, $\delta>0$, and let $Z^n(m)$, $m\in [1: 2^{nR}]$, be conditionally independent random sequences distributed according to $\prod_{i=1}^n p_{Z}(z_i)$. Suppose that the sequence $X^n$ is pairwise independent of the sequences $Z^n(m)$, $m\in [1:2^{nR}]$. Then,
\begin{align}%
\prob{ (Z^n(m),X^n)\notin\tset(p_{Z,X}) \,\text{for all $m\in [1: 2^{nR}]$}  }\leq%
\exp\left( -2^{n(R- I(Z;X)-\eps_n(\delta )} \right)
\end{align}%
where $\eps_n(\delta)$ tends to zero as $n\rightarrow\infty$ and $\delta\rightarrow 0$.
\end{lemma}
Let $X^n\sim \prod_{i=1}^n p_X(x_i)$ be an information source sequence,  encoded by an index $m$ at compression rate $R$.
Based on the covering lemma above, as long as the compression rate is higher than $I(Z;X)$,
a set of random codewords, $ Z^n(m)\sim \prod_{i=1}^n p_Z(z_i) $, contains with high probability at least one sequence that is jointly typical with the source sequence.

Though originally stated in the context of lossy source coding, the classical covering lemma is useful in a variety of scenarios
\cite{ElGamalKim:11b}, including the random-parameter channel with non-causal CSI. In this case, the parameter sequence 
$S^n\sim \prod_{i=1}^n q(s_i)$ plays the role of the ``source sequence".

\subsection{Quantum Typical Subspaces}
Moving to the quantum method of types, 
suppose that the state of a system is generated from an ensemble $\{ p_X(x), |x\rangle \}_{x\in\Xset}$, hence, the average density operator is
\begin{align}
\rho=\sum_{x\in\Xset} p_X(x) \kb{x} \,.
\end{align}
Consider the  subspace  spanned by the vectors $| x^n \rangle$, $x^n\in\Tset(\hP)$, for a given type
$\hP\in\pSpace_n(\Xset)$. %
Then, %
the projector onto the subspace %
is given by
\begin{align}
\Pi_{A^n}(\hP)\equiv \sum_{x^n\in\Tset(\hP)} \kb{ x^n } \,.
\label{eq:ProjType}
\end{align}
Note that the dimension of  the subspace of type class $\hP$ is given by %
$\trace(\Pi_{A^n}(\hP))=|\Tset(\hP)|$. By classical type properties \cite[Lemma 2.3]{CsiszarKorner:82b} (see also  \cite[Property 15.3.2]{Wilde:17b}),
\begin{align}
(n+1)^{|\Xset|} 2^{nH(\rho)} \leq    \trace(\Pi_{A^n}(\hP))  \leq  2^{nH(\rho)} \,.
\label{eq:Tsize}
\end{align}

The projector onto the $\delta$-typical subspace is defined as
\begin{align}
\Pi^\delta(\rho)\equiv \sum_{x^n\in\tset(p_X)} \kb{ x^n } \,.
\end{align}
Based on \cite{Schumacher:95p} \cite[Theorem 12.5]{NielsenChuang:02b}, for every $\eps,\delta>0$ and sufficiently large $n$, the $\delta$-typical projector satisfies
\begin{align}
\trace( \Pi^\delta(\rho) \rho^{\otimes n} )\geq& 1-\eps  \label{eq:UnitT} \\
 2^{-n(H(\rho)+c\delta)} \Pi^\delta(\rho) \preceq& \,\Pi^\delta(\rho) \,\rho^{\otimes n}\, \Pi^\delta(\rho) \,
\preceq 2^{-n(H(\rho)-c\delta)}\, \Pi^\delta(\rho)
\label{eq:rhonProjIneq}
\\
\trace( \Pi^\delta(\rho))\leq& 2^{n(H(\rho)+c\delta)} \label{eq:Pidim}
\end{align}
where $c>0$ is a constant.

We will also need the conditional $\delta$-typical subspace. Consider a  state
\begin{align}
\sigma=\sum_{x\in\Yset} p_{X}(x) \rho_B^x 
\end{align}
with 
\begin{align}
 \rho_B^x =\sum_{y\in\Yset} p_{Y|X}(y|x) \kb{\psi^{x,y}} \,.
\end{align}
Given a fixed sequence $x^n\in\Xset^n$, divide the index set $[1:n]$ into the subsets $I_n(a)=\{ i: x_i=a  \}$, $a\in\Xset$,
and define the conditional $\delta$-typical subspace $\mathscr{S}^\delta(\sigma_B|x^n)$ as the span of the vectors
$|\psi^{x^n,y^n}\rangle=\otimes_{i=1}^n |\psi^{x_i,y_i}\rangle$ such that %
\begin{align}
y^{I_n(a)}\in \Aset_\delta^{(|I_n(a)|)}(p_{Y|X=a}) \,,\;\text{for $a\in\Xset$}\,.
\end{align}
The projector onto the conditional $\delta$-typical subspace is defined as
\begin{align}
\Pi^\delta(\sigma_B|x^n)\equiv %
\sum_{|\psi^{x^n,y^n}\rangle\in\mathscr{S}^\delta(\sigma_B|x^n)} \kb{ \psi^{x^n,y^n} }
 \,.
\end{align}
Based on \cite{Schumacher:95p} \cite[Section 15.2.4]{Wilde:17b}, for every $\eps',\delta>0$ and sufficiently large $n$, %
\begin{align}
\trace( \Pi^\delta(\sigma_B|x^n) \rho_{B^n}^{x^ n} )\geq& 1-\eps'  \label{eq:UnitTCond} \\
 2^{-n(H(B|X')_\sigma+c'\delta)} \Pi^\delta(\sigma_B|x^n) \preceq& \,\Pi^\delta(\sigma_B|x^n) \,\rho_{B^n}^{x^ n}\, \Pi^\delta(\sigma_B|x^n) \,
\preceq 2^{-n(H(B|X')_{\sigma}-c'\delta)}\Pi^\delta(\sigma_B|x^n)
\label{eq:rhonProjIneqCond}
\\
\trace( \Pi^\delta(\sigma_B|x^n))\leq& 2^{n(H(B|X')_\sigma+c'\delta)} \label{eq:PidimCond}
\end{align}
where $c'>0$ is a constant, $\rho_{B^n}^{x^n}=\bigotimes_{i=1}^n \rho_{B_i}^{x_i}$, and the classical random variable $X'$ is distributed according to the type of $x^n$.
Furthermore, if $x^n\in\tset(p_X)$, then %
\begin{align}
\trace( \Pi^\delta(\sigma_B) \rho_{B^n}^{x^n} )\geq& 1-\eps' \,. 
\label{eq:UnitTCondB}
\end{align}
 (see \cite[Property 15.2.7]{Wilde:17b}).
We note that the conditional entropy in the bounds above can also be expressed as 
\begin{align}
H(B|X')_\sigma=\frac{1}{n} H(B^n|X^n=x^n)_{\sigma}  \equiv \frac{1}{n} H(B^n)_{\rho^{x^n}} \,.
\end{align}

\subsection{Quantum Packing Lemma}
To prove achievability for the HSW Theorem (see Theorem~\ref{theo:CeaNoSI}), one may invoke the quantum packing lemma \cite{HsiehDevetakWinter:08p,Wilde:17b}.
Suppose that Alice employs a codebook that consists  of $2^{nR}$ codewords $x^n(m)$, $m\in [1:2^{nR}]$, by which she chooses a quantum state from an ensemble $\{\rho_{x^n} \}_{x^n\in\Xset^n}$. The proof is based on random codebook generation, where the codewords are drawn at random according to an input distribution $p_X(x)$. To recover the transmitted message, Bob may perform the square-root measurement \cite{Holevo:98p,SchumacherWestmoreland:97p} using
a code projector $\Pi$ and codeword projectors $\Pi_{x^n}$, $x^n\in\Xset^n$, which project onto subspaces of the Hilbert space 
$\Hset_{B^n}$. 
The lemma below is a  simplified, less general, version of the quantum packing lemma by Hsieh, Devetak, and Winter \cite{HsiehDevetakWinter:08p}.
\begin{lemma}[Quantum Packing Lemma {\cite[Lemma 2]{HsiehDevetakWinter:08p}}]
\label{lemm:Qpacking}
Let %
\begin{align}
\rho=\sum_{x\in\Xset} p_X(x) \rho_x
\end{align}
where $\{ p_X(x), \rho_x \}_{x\in\Xset}$ is a given ensemble. %
Furthermore, suppose that there is  a code projector $\Pi$ and codeword projectors $\Pi_{x^n}$, $x^n\in\tset(p_X)$, that satisfy for every 
$\alpha>0$ and sufficiently large $n$,
\begin{align}
\trace(\Pi\rho_{x^n})\geq&\, 1-\alpha \\
\trace(\Pi_{x^n}\rho_{x^n})\geq&\, 1-\alpha \\
\trace(\Pi_{x^n})\leq&\, 2^{n e_0}\\
\Pi \rho^{\otimes n} \Pi \preceq&\, 2^{-n(E_0-\alpha)} \Pi 
\end{align}
for some $0<e_0<E_0$ with $\rho_{x^n}\equiv \bigotimes_{i=1}^n \rho_{x_i}$.
Then, there exist codewords $x^n(m)$, $m\in [1:2^{nR}]$, and  a POVM $\{ \Lambda_m \}_{m\in [1:2^{nR}]}$ such that 
\begin{align}
\label{eq:QpackB}
  \trace\left( \Lambda_m \rho_{x^n(m)} \right)  \geq 1-2^{-n[ E_0-e_0-R-\eps_n(\alpha)]}
\end{align}
for all %
$m\in [1:2^{nR}]$, where $\eps_n(\alpha)$ tends to zero as $n\rightarrow\infty$ and $\alpha\rightarrow 0$. 
\end{lemma}
In our analysis, where there is CSI at the encoder, we  apply the packing lemma such that the quantum ensemble encodes both the message $m$ and a compressed representation of the parameter sequence $s^n$.

\subsection{Gentle Measurement}
The gentle measurement lemma is a useful tool. As will be seen, it guarantees that we can perform multiple measurements such that the 
state of the system remains almost the same after each measurement.

\begin{lemma}[see {\cite{Winter:99p,OgawaNagaoka:07p}}]
\label{lemm:gentleM}
Let $\rho$ be a density operator. Suppose that $\Lambda$ is a meaurement operator such that $0\preceq \Lambda\preceq \identity$. If
\begin{align}
\trace(\Lambda\rho) \geq 1-\eps
\end{align}
for some $0\leq\eps\leq 1$, then the post-measurement state $\rho'\equiv \frac{\sqrt{\Lambda}\rho\sqrt{\Lambda} }{\trace(\Lambda\rho)}$ is $2\sqrt{\eps}$-close to the original state in trace distance, \ie
\begin{align}
\norm{ \rho-\rho' }_1\leq 2\sqrt{\eps} \,.
\end{align}
\end{lemma}
The lemma is particularly useful in our analysis since the POVM operators in the quantum packing lemma satisfy the conditions of the lemma for large $n$ (see (\ref{eq:QpackB})).

\section{Proof of Lemma~\ref{lemm:pureCeaSCausal}}
\label{app:pureCeaSCausal}
Consider the region $\mathcal{R}_{\text{s-c}}(\channel)$ as defined in (\ref{eq:inCeaSCausal}).

\subsection{Purification}
To prove that a union over pure states is sufficient, we show that for every rate $R_0$ that can be achieved with distortion $D$, there exists a rate $R_1\geq R_0$  that can be achieved with pure states and the same distortion.
Fix $p_X(x)p_{Z|X,S}(z|x,s)$, $\{ \theta^{x}_{A} \}$, and  $\{ \Gamma_{B|x,z}^{\hs} \}$. Let  
\begin{align}
R_0=& I(X,Z;B)_\rho-I(Z;S|X) %
\label{eq:scR0p}
\\
D_0=& \sum_{s\in\Sset}  \sum_{x\in\Xset} \sum_{z\in\Zset} \sum_{\hs\in\widehat{S}} q(s) p_X(x)p_{Z|X,S}(z|x,s)   \trace(  \Gamma_{B|x,z}^{\hs} \rho^{s,x}_{B} ) d(s,\hs)
\label{eq:scD0p}
\end{align}
and consider the spectral decomposition,
\begin{align}
\theta^{x}_{A}=\sum_{w\in\Wset} p_{W|X}(w|x) \phi^{x,w}_A 
\label{eq:scPureThetaxz}
\end{align}
where $P_{W|X}(w|x)$ is a conditional  probability distribution, and $\phi^{x,w}_A$ are pure. 
Consider the extended state
\begin{multline}
\rho_{SXZWA}=\sum_{s,x,z,w} q(s)p_{X}(x)p_{Z|X,S}(z|x,s)p_{W|X}(w|x) \kb{s} \otimes \kb{x}%
 \otimes \kb{z} \otimes \kb{w}
\otimes \phi^{x,w}_A \,.
\label{eq:scPureSXZWA}
\end{multline}

Now, observe that the union in the RHS of (\ref{eq:inCeaSCausal}) includes
the rate-distortion pair $(R_1,D_1)$ that is given by
\begin{align}
R_1=& I(X,W,Z;B)_\rho-I(Z;S|X,W) %
\label{eq:scR1p}
\\
D_1=& \sum_{s,x,z,w,\hs} q(s) p_X(x)p_{Z|X,S}(z|x,s) p_{W|X}(w|x)   \trace\left(  \Gamma_{B|x,z}^{\hs}
\channel^{(s)}( \phi^{x,w}_{A} ) \right) d(s,\hs)
\label{eq:scD1p}
\end{align}
which is obtained by plugging $X'=(X,W)$ instead of $X$, and the pure states $\phi^{x,w}_A$ instead of $\theta_A^{x}$.
That is, $(R_1,D_1)\in\mathcal{R}_{\text{s-c}}(\channel)$. 
 According to (\ref{eq:scPureSXZWA}), the random variables $S\Cbar (X,Z)\Cbar W$ form a Markov chain, thus $I(W;S|X,Z)=0$. 
By the chain rule, it follows that $I(Z;S|X,W)=I(Z;S|X)+I(W;S|X,Z)-I(W;S|X)=I(Z;S|X)$, hence
$R_1=I(X,W,Z;B)_\rho-I(Z;S|X)\geq I(X,Z;B)_\rho-I(Z;S|X)=R_0 $. As for the distortion level, we have by linearity that
\begin{align}
D_1&=\sum_{s,x,z,\hs} q(s) p_X(x)p_{Z|X,S}(z|x,s)    \trace\left(  \Gamma_{B|x,z}^{\hs}
\channel^{(s)}\left( \sum_w p_{W|X}(w|x) \phi^{x,w}_{A} \right)\right) d(s,\hs) 
\nonumber\\&
=D_0
\end{align}
where the last equality is due to (\ref{eq:scD0p}) and (\ref{eq:scPureThetaxz}).
Thereby, the union can be restricted to pure states.

\subsection{Cardinality Bounds}
To bound the alphabet size of the random variables $X$ and $Z$, we use the Fenchel-Eggleston-Carath\'eodory lemma \cite{Eggleston:66p} and similar arguments as 
in \cite{YardHaydenDevetak:08p}.
Let
\begin{align}
L_0=&|\Hset_A|^2+1 \\
L_1=& |\Hset_A|^2+|\Sset| \,.
\end{align}

First, fix $q(s)$ and $p_{Z|X,S}(z|x,s)$, and consider the ensemble $\{ p_X(x)p_{Z|X,S}(z|x,s) \,, \theta_A^{x}  \}$. 
An Hermitian matrix can be specified by  $|\Hset_A|$ real values for the diagonal and $\frac{1}{2}|\Hset_A|(|\Hset|_A-1)$ complex numbers for the non-diagonal entries, or, $|\Hset_A|^2$ real parameters in total. Since a density matrix is Hermitian and also has a unit trace,
every quantum state $\theta_A$ has a unique parametric representation $u(\theta_A)$ of dimension $|\Hset_A|^2-1$. 
Then, define a map $f_0:\Xset\rightarrow \mathbb{R}^{L_0}$ by
\begin{align}
f_0(x)= \left(  u(\theta_A^x) \,,\; -H(B|X=x,Z)_{\rho}+H(S|X=x,Z) \,,\; \E[ d(S,\hS) |X=x ]   \right) \,.
\end{align}
The map $f_0$ can be extended to probability distributions as follows,
\begin{align}
F_0 \,:\; p_X  \mapsto
\sum_{x\in\Xset} p_X(x) f_0(x)= \left(  u(\theta_A) \,,\; -H(B|X,Z)_{\rho}+H(S|X,Z) \,,\; \E d(S,\hS)    \right) %
\end{align}
where $\theta_A=\sum_x p_X(x) \theta_A^x$.
According to the Fenchel-Eggleston-Carath\'eodory lemma \cite{Eggleston:66p}, any point in the convex closure of a connected compact set within $\mathbb{R}^d$ belongs to the convex hull of $d$ points in the set. 
Since the map $F_0$ is linear, it maps  the set of distributions on $\Xset$ to a connected compact set in $\mathbb{R}^{L_0}$. Thus, for every  $p_X$, 
there exists a probability distribution $p_{\bar{X}}$ on a subset $\overline{\Xset}\subseteq \Xset$ of size $%
L_0$, such that 
$%
F_0(p_{\bar{X}})=F_0(p_{X}) %
$. %
We deduce that alphabet size can be restricted to $|\Xset|\leq L_0$, while preserving $\theta_A$ and
$\rho_B\equiv \sum_s q(s)\channel^{(s)}(\theta_A)$; $I(X,Z;B)_\rho-I(Z;S|X)=H(B)_\rho-H(B|X,Z)_{\rho}+H(S|X,Z)-H(S)$; and
$\E d(S,\hS)$.

We move to the alphabet size of $Z$. Fix $p_{X,S|Z}$, where
\begin{align}
p_{X,S|Z}(x,s|z)\equiv \frac{ q(s) p_X(x) p_{Z|X,S}(z|x,s)}{ \sum_{s'\in\Sset} q(s')  \sum_{x'\in\Sset} p_X(x') p_{Z|X,S}(z|x',s')}
\,.
\end{align}
Define the map $f_1:\Zset\rightarrow \mathbb{R}^{L_1}$ by
\begin{align}
f_{1}(z)=& \left( p_{S|Z}(\cdot|z) %
\,,\; -H(B|X,Z=z)_{\rho}+H(S|X,Z=z) \,,\; \E[ d(S,\hS) |Z=z ]  \right) \,.
\end{align}
Now, the extended map is
\begin{align}
F_1 \,:\; p_{Z} \mapsto  \sum_{z\in\Zset} p_{Z}(z) f_1(z) = \left( q(\cdot) %
\,,\; -H(B|X,Z)_{\rho}+H(S|X,Z) \,,\; \E d(S,\hS)     \right) \,.
\end{align}
By the Fenchel-Eggleston-Carath\'eodory lemma \cite{Eggleston:66p}, for every  $p_{Z}$, 
there exists $p_{\bar{Z}}$ on a subset $\overline{\Zset}\subseteq \Zset$ of size $ L_1 $, such that 
$F_1(p_{\bar{Z}})=F_1(p_Z)$.
We deduce that alphabet size can be restricted to $|\Zset|\leq L_1 $, while preserving $q(s)$, %
$\rho_B$, $I(X,Z;B)_\rho-I(Z;S|X)$, and $\E d(S,\hS)$.
\qed

\section{Proof of Theorem~\ref{theo:mainSC}}
\label{app:mainSC}
Consider a random-parameter quantum channel $\channel_{SA\rightarrow B}$ with strictly-causal CSI.
\subsection*{Part 1}
\subsection{Achievability Proof}
We show that for every $\zeta_0,\eps_0,\delta_0>0$, there exists a $(2^{n(R-\zeta_0)},n,\eps_0,D+\delta_0)$ code for $\channel_{SA\rightarrow B}$ with strictly-causal CSI, provided that $(R,D)\in \mathcal{R}_{\text{s-c}}(\channel)$. 
To prove achievability, we extend  the classical block Markov coding to the quantum setting, and then apply the quantum packing lemma and the classical covering lemma. We use the gentle measurement lemma \cite{Winter:99p}, %
which guarantees that multiple decoding measurements can be performed without ``destroying" the output state.

Recall that with strictly-causal CSI, the encoder has access to the sequence of \emph{past} parameters $s_1,s_2,\ldots,s_{i-1}$.
Let $\{ p_X(x)p_{Z|X,S}(z|x,s) , \theta_{A}^{x} \}$  be a  given ensemble, and fix a set of POVMs $\{ \Gamma_{B|x,z}^{\hs} \}$ such that
\begin{align}
 \sum\limits_{s,\hs,x,z} q(s)p_X(x) p_{Z|X,S}(z|x,s) %
						\trace( \Gamma_{B|x,z}^{\hs} \channel^{(s)}(\theta^{x}_{A}) ) d(s,\hs) \leq D \,.
\label{eq:ICscDist}
\end{align}
Define the average states
\begin{align}
\rho_B^x=& \sum_{s\in\Sset} q(s) \channel^{(s)} (\theta_A^{x})%
\,,\;\text{for $x\in\Xset$,}
\\
\rho_B=& \sum_{x\in\Xset} p_X(x) \rho_B^x %
\end{align}
We will also consider the a posteriori probability distribution, conditioning on $Z=z$:
\begin{align}
\hat{p}_{X,S|Z}(x,s|z)=\frac{q(s)p_X(x) p_{Z|X,S}(z|x,s)}{\sum_{x'\in\Xset}\sum_{s'\in\Sset} q(s')p_X(x') p_{Z|X,S}(z|x',s')} \,.
\end{align}
Then, the corresponding output state is
\begin{align}
\sigma_B^{x,z}=\sum_{s\in\Sset} \hat{p}_{S|Z,X}(s|z,x) \channel^{(s)} (\theta_A^{x}) \,.
\end{align}

We use $T$ transmission blocks, where each block consists of $n$ input systems. In particular, with strictly-causal CSI, the encoder has access to the parameter sequences from the previous blocks.
In effect, the $j^{\text{th}}$ transmission block encodes a message $m_j\in [1:2^{nR}]$ and a compression of the parameter sequence $s^n_{j-1}$ from the previous block, 
for $j\in [2:T]$.

\vspace{-1cm}
\subsection*{ }

The code construction, encoding and decoding procedures are described below.

\subsubsection{Classical Code Construction}
Let $\delta>0$, $R_s>0$, and $\tR_s>0$ such that $R_s<\tR_s$. 
For every $j\in [2:T]$, select $2^{n(R+R_s)}$ independent sequences $x^n_j(m_j,\ell_{j-1})$, $m_j\in [1:2^{nR}]$, $\ell_{j-1}\in [1:2^{nR_s}]$,
at random according to $\prod_{i=1}^n p_X(x_{j,i})$. For every $m_j\in [1:2^{nR}]$ and $\ell_{j-1}\in [1:2^{nR_s}]$, select $2^{n \tR_s}$ conditionally independent sequences $z^n_j(k_j|m_j,\ell_{j-1})$, $k_j\in [1:2^{n \tR_s}]$,
at random according to $\prod_{i=1}^n p_{Z|X}(z_{j,i}|x_{j,i}(m_j,\ell_{j-1}))$. For $j=1$, set $\ell_0\equiv 1$, and select
$x^n_1(m_1,1)$ and $z^n_1(k_1|m_1,1)$ in the same manner, for $(m_1,k_1)\in [1:2^{nR}] \times [1:2^{n \tR_s}]$. 
 We have thus defined the classical codebooks
\begin{align}
\mathscr{B}(j)=\{ (x^n_j(m_j,\ell_{j-1}), z^n_j(k_j|m_j,\ell_{j-1}))\} \,,\; j\in [1:T]
\end{align}
with $m_j\in [1:2^{nR}]$, $\ell_{j-1}\in [1:2^{nR_s}]$, $k_j\in [1:2^{n \tR_s}]$.
Partition the set of indices $[1:2^{n \tR_s}]$ into bins $\Kset(\ell_j)=[(\ell_j-1)2^{n(\tR_s-R_s)}+1:
\ell_j 2^{n(\tR_s-R_s)}]$ of equal size $2^{n(\tR_s-R_s)}$.

\subsubsection{Encoding and Decoding}
To send the messages $(m_j)$, given the parameter sequences $(s_{1}^n,\ldots,s_{j-1}^n)$, Alice performs the following.
\begin{enumerate}[(i)]
\item
At the end of block $j$, find an index $k_j\in [1:2^{n \tR_s}]$ such that $ (s_j^n, z_j^n(k_j|m_j,\ell_{j-1}),$ $x^n_j(m_j,\ell_{j-1}))\in \tset(p_{S,X,Z}) $, where
$p_{S,X,Z}(s,x,z)=q(s)p_X(x) p_{Z|X,S}(z|x,s)$. If there is none,  select $k_j$ arbitrarily, and if there is more than one such index, choose the smallest.
Set $\ell_j$ to be the bin index of $k_j$, \ie such that $k_j\in\Kset(\ell_j)$.

\item
In block $j+1$, prepare $\rho_{A_{j+1}^n}= \bigotimes_{i=1}^n \theta_A^{x_{j+1,i}(m_{j+1},\ell_j)  }$ and send the block $A_{j+1}^n$.
\end{enumerate}
Bob receives the systems $B_1^n,\ldots,B_T^n$ in the state 
\begin{align}
\rho_{B^{Tn}}= \bigotimes_{j=1}^T \bigotimes_{i=1}^n \rho_B^{x_{j+1,i}(m_{j+1},\ell_j)}
\label{eq:rhoBTnSC}
\end{align}
 and decodes as follows. 
\begin{enumerate}[(i)]
\item
At the end of block $j+1$, decode $(\hm_{j+1},\hat{\ell}_j)$ by applying a POVM\\ $\{ \Lambda^1_{m_{j+1},\ell_j} \}_{
(m_{j+1}, \ell_j) \in  [1:2^{nR}] \times [1:2^{nR_s}]}$, which will be specified later, to the systems $B_{j+1}^n$,
for $j=0,1,\ldots,T-1$. 
\item
Decode $\hat{k}_j$ by applying a second POVM $\{ \Lambda^2_{k_j|x^n(\hm_{j+1},\hat{\ell}_j)} \}_{ k_j\in \Kset(\hat{\ell}_j) }$, which will also be specified later, to the systems $B_{j}^n$. 
\item
Reconstruct the parameter sequence by applying the POVM $\Gamma^{\hs_{j,i}}_{B|x_{j,i},z_{j,i}}$ to the system $B_{j,i}$ with
$x_{j,i}\equiv x_{j,i}(\hm_{j},\hat{\ell}_{j-1})$ and $z_{j,i}\equiv z_{j,i}(\hat{k}_j|\hm_j,\hat{\ell}_{j-1})$, for $j\in [1:T]$ and $i\in [1:n]$.
\end{enumerate}

\subsubsection{Analysis of Probability of Error and Distortion}
By symmetry, we may assume without loss of generality that Alice sends the message $M_j=1$ using $L_j=L_{j-1}=1$, for $j\in [1:T]$.
Consider the following events,
\begin{align}
\mathscr{E}_1(j)=& \{  (S^n,X^n(1,1),Z^n(k_j|1,1))\notin \Aset^{\delta_1}(p_{S,X,Z})  \,,\;\text{for all $k_j\in [1:2^{n\tR_s}] $} \} \\
\mathscr{E}_2(j)=& \{  (\hM_j,\hL_{j-1})\neq (1,1)  \}\\
\mathscr{E}_3(j)=& \{  \hK_j \neq K_j  \}\\
\mathscr{E}_4(j)=& \{  d^n(S_j^n,\hat{S}_j^n)>D+\frac{1}{2}\delta_0 \} 
\end{align}
with $\delta_1\equiv \delta/(2|\Sset| |\Zset|)$.
By the union of events bound, the probability of error is bounded by
\begin{align}
P_{e|m=1}^{(Tn)}(\rho_{A^{Tn}},\Lambda_{B^{Tn}}) \leq& %
\sum_{j=1}^{T} \prob{ \mathscr{E}_1(j) }%
+\sum_{j=0}^{T-1} \cprob{ \mathscr{E}_2(j+1) }{ \mathscr{E}_1^c(j)\cap \mathscr{E}_1^c(j+1) } \nonumber\\&
+\sum_{j=0}^{T-1}\cprob{ \mathscr{E}_3(j+1) }{ \mathscr{E}_1^c(j)\cap \mathscr{E}_1^c(j+1) \cap \mathscr{E}_2^c(j+1) } \nonumber\\&
+\sum_{j=0}^{T-1}\cprob{ \mathscr{E}_4(j+1) }{ \mathscr{E}_1^c(j)\cap \mathscr{E}_1^c(j+1) \cap \mathscr{E}_2^c(j+1) \cap \mathscr{E}_3^c(j+1) } 
\label{eq:PeBsc}
\end{align}
where the conditioning on $M_j=L_j=L_{j-1}=1$ is omitted for convenience of notation.
By the classical covering lemma, the probability terms $\prob{ \mathscr{E}_1(j) }$ tend to zero as $n\rightarrow\infty$ for
\begin{align}
\tR_s> I(X,Z;S)+\eps_1(\delta)=I(Z;S|X)+\eps_1(\delta) %
\label{eq:B1}
\end{align}
where the last equality holds since the random variables $X$ and $S$ are statistically independent,
using the notation $\eps_i(\delta)$ %
for terms that tend to zero as $\delta\rightarrow 0$.

To bound the second sum, we use the quantum packing lemma. Given $\mathscr{E}_1^c(j)$, we have that $X^n(1,1)\in\Aset^{
\nicefrac{\delta}{2}}(p_X)$. %
Now, observe that
\begin{align}
\Pi^{\delta}(\rho_B)  \rho_{B^n}   \Pi^{\delta}(\rho_B) \preceq& 2^{ -n(H(B)_{\rho}-\eps_2(\delta)) } \Pi^{\delta}(\rho_B)
\\
\trace\left[ \Pi^{\delta}(\rho_B|x^n) \rho_{B^n}^{x^n} \right] \geq& 1-\eps_2(\delta) \\
\trace\left[ \Pi^{\delta}(\rho_B|x^n)  \right] \leq& 2^{ n(H(B|X)_{\rho} +\eps_2(\delta))} \\
\trace\left[ \Pi^{\delta}(\rho_B) \rho_{B^n}^{x^n} \right] \geq& 1-\eps_2(\delta) 
\end{align}
for all $x^n\in\Aset^{\delta_1}(p_X)$, by (\ref{eq:rhonProjIneq}), (\ref{eq:UnitTCond}),  (\ref{eq:PidimCond}), and 
(\ref{eq:UnitTCondB}), respectively. Since the codebooks are statistically independent of each other, we have by Lemma~\ref{lemm:Qpacking} that there exists a POVM $\Lambda^1_{m_{j+1},\ell_j}$ such that
$%
\cprob{ \mathscr{E}_2(j+1) }{ \mathscr{E}_1^c(j)\cap \mathscr{E}_1^c(j+1) } \leq 2^{ -n( I(X;B)_\rho -(R+R_s)-\eps_3(\delta)) } 
$, %
which tends to zero as $n\rightarrow\infty$, provided that 
\begin{align}
R< I(X;B)_\rho -R_s-\eps_3(\delta) \,.
\label{eq:B2}
\end{align}

Moving to the third sum in the RHS of (\ref{eq:PeBsc}), suppose that $\mathscr{E}_2^c(j+1)$ occurred, namely the decoder measured the correct $M_{j+1}$ and $L_j$. Denote the state of the systems $B_j^n$ after this measurement by $\rho'_{B_j^n}$.
Then, observe that due to the packing lemma inequality (\ref{eq:QpackB}),  Lemma~\ref{lemm:gentleM} (the gentle measurement lemma) implies that the post-measurement state is close to the original state in the sense that
\begin{align}
\frac{1}{2}\norm{\rho'_{B_j^n}-\rho_{B_j^n}}_1 \leq 2^{ -n\frac{1}{2}( I(X;B)_\rho -(R+R_s)-\eps_4(\delta)) } \leq \eps_5(\delta)
\end{align}
for sufficiently large $n$ and rates as in (\ref{eq:B2}).
Therefore, the distribution of measurement outcomes when $\rho'_{B_j^n}$ is measured is roughly the same as if the POVM $\Lambda^1_{m_{j+1},\ell_j}$ was never performed. To be precise, the difference between the probability of a measurement outcome $\widehat{k}_j$ when $\rho'_{B_j^n}$ is measured and the probability when $\rho_{B_j^n}$ is measured is bounded by $ \eps_5(\delta)$ in absolute value \cite[Lemma 9.11]{Wilde:17b}.
Furthermore, 
\begin{align}
\trace\left[ \Pi^{\delta}(\rho_B|x^n,z^n) \sigma_{B^n}^{x^n,z^n} \right] \geq& 1-\eps_6(\delta) \\
\Pi^{\delta}(\rho_B|x^n)  \rho_{B^n}^{x^n}   \Pi^{\delta}(\rho_B|x^n) \preceq& 2^{ -n(H(B|X)_{\rho}-\eps_6(\delta)) } \Pi^{\delta}(\rho_B|x^n)
\\
\trace\left[ \Pi^{\delta}(\rho_B|x^n,z^n)  \right] \leq& 2^{ n(H(B|X,Z)_{\rho} +\eps_6(\delta))} \\
\trace\left[ \Pi^{\delta}(\rho_B|x^n) \sigma_{B^n}^{x^n,z^n} \right] \geq& 1-\eps_6(\delta) 
\end{align}
for all $(x^n,z^n)\in\Aset^{\nicefrac{\delta}{2}}(p_X p_{Z|X})$, by (\ref{eq:UnitTCond}), (\ref{eq:rhonProjIneqCond}), (\ref{eq:PidimCond}), and 
(\ref{eq:UnitTCondB}), respectively. Therefore, we have by the packing lemma that there exists a POVM $\Lambda^2_{k_{j}|x^n}$ such that
\begin{align}
\cprob{ \mathscr{E}_3(j+1) }{ \mathscr{E}_1^c(j)\cap \mathscr{E}_1^c(j+1) \cap \mathscr{E}_2^c(j+1) } \leq 2^{ -n( I(Z;B|X)_\rho -(\tR_s-R_s)-\eps_7(\delta)) } 
\end{align}
which tends to zero as $n\rightarrow\infty$, provided that
\begin{align}
R_s> \tR_s- I(Z;B|X)_\rho +\eps_7(\delta) \,.
\label{eq:B3}
\end{align}

It remains to verify that the distortion requirement is satisfied. Suppose that $\mathscr{E}_3^c(j+1)$ occurred, namely the decoder measured the correct $K_{j+1}$. Denote the post-measurement state by $\rho''_{B_j^n}$.
As before, the gentle measurement lemma guarantees that the difference between the probability of a measurement outcome $\widehat{s}$ when $\rho''_{B_j^n}$ is measured and the probability when $\rho'_{B_j^n}$ is measured is bounded by $ \eps_5(\delta)$ in absolute value.
 Therefore, given $\mathscr{E}_1^c(j)\cap \mathscr{E}_1^c(j+1) \cap \mathscr{E}_2^c(j+1)\cap \mathscr{E}_3^c(j+1)$,  the parameter sequence $S_{j+1}^n$ and the reconstruction $\hat{S}_{j+1}^n$ have a product distribution that is $2\eps_5(\delta)$-close to 
\begin{align}
\prob{ S=s,\hat{S}=\hs }= q(s)  \sum_{x,z} p_X(x) p_{Z|X}(z|x) \trace( \Gamma^{\hs}_{B|x,z} \channel^{(s)} ( \theta_A^{x} ) ) \,.
\end{align}
By (\ref{eq:ICscDist}), the distribution above satisfies $\E d(S,\hat{S})\leq  D$, hence the last term tends to zero as $n\rightarrow\infty$ by the law of large numbers. By the law of total expectation, %
\begin{align}
\E d^{Tn}(S^{Tn},\hS^{Tn})\leq 
\sum_{j=1}^T \prob{\mathscr{E}_1(j)\cup\mathscr{E}_2(j)\cup\mathscr{E}_3(j)\cup\mathscr{E}_4(j)} d_{\max} + D+\frac{1}{2}\delta_0 \,. %
\end{align}
 Thereby, the asymptotic average distortion is bounded by $(D+\delta_0)$ and the probability of error tends to zero as $n\rightarrow\infty$ for rates that satisfy (\ref{eq:B1}), (\ref{eq:B2}), and (\ref{eq:B3}), which requires 
\begin{align}
R<& I(X;B)_\rho -(I(Z;S|X)-I(Z;B|X)_\rho+\eps_1(\delta)+\eps_7(\delta))-\eps_3(\delta)  \nonumber\\ =&
I(X,Z;B)_\rho -I(Z;S|X)-\eps_8(\delta) \,.
\end{align}
To show that rate-distortion pairs in $\frac{1}{\kappa}\mathcal{R}_{\text{s-c}}(\channel^{\otimes \kappa})$ are achievable as well, one may employ the coding scheme above for the product channel $\channel^{\otimes \kappa}$, where $\kappa$ is arbitrarily large.
This completes the proof of the direct part.

\subsection{Converse Proof}
Consider the converse part for the regularized capacity formula. As can be seen below, a regularized converse is straightforward.
Let $M$ be a uniformly distributed message.
 Suppose that at time $i\in [1:n]$, Alice  sends 
$\rho^{m,s^{i-1}}_{ A_i}$ over the channel. 
After Alice sends the systems $A^n$ through the channel, Bob receives the systems $B^n$ in the state
$\rho_{B^n}=\frac{1}{2^{nR}}\sum_{m=1}^{2^{nR}} \sum_{s^n\in\Sset^n} q^n(s^n)\channel^{(s^n)}(\overline{\Fset}(m,s^n)) $
with
\begin{align}
\overline{\Fset}_{M, S^n\rightarrow A^n}= \bigotimes_{i=1}^n \Fset^{(i)}_{M,S^{i-1}\rightarrow A_i} \,.
\end{align}
Then, Bob performs a decoding POVM $\Lambda^{m,\hs^n}_{B^n}$.

Consider a sequence of codes $(\overline{\Fset}_{MS^n\rightarrow A^n},\Lambda_{B^n})$ such that the average probability of error tends to zero and the distortion requirement holds. That is,
\begin{align}
&%
\prob{ \hM\neq M } \leq \alpha_n \,,
\label{eq:convPe1}
\intertext{and}
& \Delta^{(n)} (\Fset,\Lambda)  \leq D \,.
\end{align}
By Fano's inequality, (\ref{eq:convPe1}) implies that
$%
H(M|\hM) \leq n\eps_n
$, %
where $\eps_n$ tends to zero as $n\rightarrow\infty$.
Hence, 
\begin{align}
nR= H(M) \leq& I(M;\hM)_{\rho}+n\eps_n %
\leq I(M;B^n)_{\rho}+n\eps_n 
\label{eq:ConvIneq1SC}
\end{align}
where the last inequality follows from the Holevo bound \cite[Theorem 12.1]{NielsenChuang:02b}.
Since $M$ and $S^n$ are statistically independent, we can write the last bound as
\begin{align}
R\leq \frac{1}{n} [ I(M;B^n)_{\rho}-I(M;S^n)  ]+ \eps_n = \frac{1}{n}   [ I(X^n,Z^n;B^n)_{\rho}-I(X^n,Z^n;S^n)  ]+ \eps_n
\end{align}
for $X^n=f(M)$ and $Z^n=\emptyset$, where $f$ is an arbitrary one-to-one map from $[1:2^{nR}]$ to $\Xset^n$. 

As for the distortion requirement,
\begin{align}
D\geq& 
 \Delta^{(n)}(\Eset,\Lambda)= \E d^n(S^n,\hS^n) \nonumber\\=& 
P_e^{(n)} (\rho_{A^n},\Lambda_{B^n}) \E[ d^n(S^n,\hS^n) \,|\; \hM\neq M ] + (1- P_e^{(n)} (\rho_{A^n},\Lambda_{B^n})) \E[ d^n(S^n,\hS^n) \,|\; \hM= M ]
\nonumber\\
\geq&  (1- P_e^{(n)} (\rho_{A^n},\Lambda_{B^n})) \E[ d^n(S^n,\hS^n) \,|\; \hM= M ]
\nonumber\\
\geq& (1-\alpha_n)  \E[ d^n(S^n,\hS^n) \,|\; \hM= M ]
\end{align}
where we have used the law of total expectation in the second line, and (\ref{eq:convPe1}) in the last line. 
Thus,  
\begin{align}
D\geq&   \E[ d^n(S^n,\hS^n) \,|\; \hM= M ] -\alpha_n d_{\max} 
\nonumber\\
=& \sum_{s^n\in\Sset^n} \sum_{\hs^n\in\widehat{\Sset}^n}
d^n(s^n,\hs^n) q^n(s^n) \cdot \frac{1}{2^{nR}} \sum_{m=1}^{2^{nR}} \trace\Big[ \Lambda^{m,\hat{s}^n}_{B^n}
\channel^{(s^n)}_{A^n\rightarrow B^n}
(\overline{\rho}_{A^n}^{m,s^n}) \Big] -\alpha_n d_{\max}
\label{eq:distortionSCconvN}  \\ =&
 \sum_{s^n\in\Sset^n}  \sum_{x^n\in\Xset^n} %
\sum_{\hs^n\in\widehat{S}^n} q^n(s^n) p_{X^n}(x^n)%
 \trace(  \Gamma_{B^n|x^n}^{\hs^n} \rho^{s^n,x^n}_{B^n} ) d^n(s^n,\hs^n)
\end{align}
with $\Gamma_{B^n|f(m)}^{\hs^n}\equiv \Lambda^{m,\hat{s}^n}_{B^n}$.
This concludes the converse proof for part 1.

\subsection*{Part 2}
Now, we consider the quantum-classical special case of a measurement channel $\Mset_{SA\rightarrow Y}$. The direct part follows from 
part 1, taking $\kappa=1$.
It remains to prove the converse part, which we show by extending the methods of Choudhuri \etal \cite{ChoudhuriKimMitra:13p}. 

 By (\ref{eq:ConvIneq1SC}) and the chain rule for classical mutual information, we have
\begin{align}
nR\leq& I(M;Y^n)+n\eps_n= \sum_{i=1}^n I(M;Y_i|Y_{i+1}^n) +n\eps_n \,.
\label{eq:ConvIneq1SCm}
\end{align}
We can rewrite the bound above as
\begin{align}
R-\eps_n\leq& \frac{1}{n}\sum_{i=1}^n [ I(M,S^{i-1};Y_i|Y_{i+1}^n)-I(S^{i-1};Y_i|M,Y_{i+1}^n)]  \nonumber\\
=&  \frac{1}{n}\sum_{i=1}^n [ I(M,S^{i-1};Y_i|Y_{i+1}^n)-I(Y_{i+1}^n;S_i|M,S^{i-1})]  \nonumber\\
\leq&  \frac{1}{n}\sum_{i=1}^n [ I(M,S^{i-1},Y_{i+1}^n;Y_i)-I(Y_{i+1}^n;S_i|M,S^{i-1})] 
\label{eq:ConvIneq2SCm}
\end{align}
where the equality follows from the Csisz\'ar sum identity \cite[Section 2.3]{ElGamalKim:11b}. Since the pair $(M,S^{i-1})$ is statistically independent of $S_i$, we have $I(Y_{i+1}^n;S_i|M,S^{i-1})=I(M,S^{i-1},$ $Y_{i+1}^n;S_i)$, hence
\begin{align}
R-\eps_n\leq \frac{1}{n}\sum_{i=1}^n [ I(X_i,Z_i;Y_i)-I(X_i,Z_i;S_i)] 
\end{align}
where we have defined $X_i=(M,S^{i-1})$ and $Z_i=Y_{i+1}^n$. Thus,
\begin{align}
R-\eps_n\leq  I(X,Z;Y|J)-I(X,Z;S|J)  
\end{align}
with 
\begin{align}
X\equiv X_J \,,\;  S=S_J \,,\;  Y=Y_J \,,\; \hS=\hS_J 
\label{eq:convSCxsyhs}
\end{align}
where $J$ is uniformly distributed over $[1:n]$, and independent of $(M,S^n)$.
Then, defining $X'=(X,J)$, we have that $I(X,Z;Y|J)\leq I(X',Z;Y)$ and $I(X,Z;S|J)=I(X',Z;S)=I(Z;S|X')$, hence 
$%
R-\eps_n\leq  I(X',Z;Y)-I(Z;S|X')  
$. %

As for the distortion level, 
\begin{align}
D\geq& \E d^n(S^n,\hS^n)= \frac{1}{n} \sum_{i=1}^n \E d(S_i,\hS_i) =\E d(S,\hS)   
\end{align}
where the first equality holds since the distortion measure is additive (see (\ref{eq:dn})), and the second follows from the definition of $S$ and $\hS$ in (\ref{eq:convSCxsyhs}).
This completes the proof of Theorem~\ref{theo:mainSC}.
\qed

\section{Proof of Theorem~\ref{theo:mainC}}
\label{app:mainC}
Since the proof is similar to that of Theorem~\ref{theo:mainSC} in Appendix~\ref{app:mainSC}, we only give an outline. 
The converse proof in part 1 follows the same arguments, and it is thus omitted. 
Moving to the achievability proof, %
we need to show that for every $\zeta_0,\eps_0,\delta_0>0$, there exists a $(2^{n(R-\zeta_0)},n,\eps_0,D+\delta_0)$ code for $\channel_{SA\rightarrow B}$ with causal CSI, provided that $(R,D)\in \mathcal{R}_{\text{caus}}(\channel)$. 
Here, the encoder has access to the sequence of past \emph{and present} parameters $s_1,s_2,\ldots,s_i$.
Let $\{ p_X(x)p_{Z|X}(z|x) , \theta_{G}^{x} \}$ be a  given ensemble, and fix the channels $\Fset^{(s)}_{G\rightarrow A}$ and set of POVMs $\{ \Gamma_{B|x,z}^{\hs} \}$ such that
\begin{align}
 \sum\limits_{s,\hs,x,z} q(s)p_X(x) p_{Z|X,S}(z|x,s) %
						\trace( \Gamma_{B|x,z}^{\hs} \Vset^{(s)}(\theta_G^{x})  )  d(s,\hs) \leq D \,.
\label{eq:ICcDist}
\end{align}
where the channel $\Vset^{(s)}_{G\rightarrow B}$ is defined by 
\begin{align}
\mathcal{V}^{(s)}(\rho_G)= \channel^{(s)}( \Fset^{(s)}(\rho_G)  ) \,.
\end{align}
Then, define the average states
\begin{align}
\rho_B^x=& \sum_{s\in\Sset} q(s) \Vset^{(s)} (\rho_G^{x}) %
\\
\rho_B=& \sum_{x\in\Xset} p_X(x) \rho_B^x 
\intertext{
and
}
\sigma_B^{x,z}=&\sum_{s\in\Sset} p_{S|X,Z}(s|x,z)  \Vset^{(s)}  ( \theta_G^{x} ) 
\end{align}
for $x\in\Xset$ and $z\in\Zset$.

We use $T$ transmission blocks, where each block consists of $n$ input systems.
The code construction, encoding and decoding procedures are described below.

\subsection*{ }
\vspace{-0.85cm}

\subsubsection*{Classical Code Construction}
The classical code construction is the same as in the previous proof:
Select i.i.d  sequences $x^n_j(m_j,\ell_{j-1})$  according to $ p_X$, and then for every $(m_j,\ell_{j-1})$, select conditionally independent sequences $z^n_j(k_j|m_j,\ell_{j-1})$ according to $p_{Z|X}$,
where $m_j\in [1:2^{nR}]$, $\ell_{j-1}\in [1:2^{nR_s}]$, $k_j\in [1:2^{n \tR_s}]$.
Partition the set of indices $[1:2^{n \tR_s}]$ into bins $\Kset(\ell_j) %
$ of equal size $2^{n(\tR_s-R_s)}$.

\subsubsection*{Encoding and Decoding}
To send the messages $(m_j)$, given the parameter sequences $(s_{1}^n,\ldots,s_j^n)$, Alice performs the following.
\begin{enumerate}[(i)]
\item
Let $\ell_0=1$.
At the end of block $j$, find an index $k_j\in [1:2^{n \tR_s}]$ such that $ (s_j^n,$ $ z_j^n(k_j|m_j,\ell_{j-1}),x^n_j(m_j,\ell_{j-1}))\in \tset(p_{S,Z,X}) $, where
$p_{S,X,Z}(s,x,z)=q(s)p_X(x)$\\ $\cdot p_{Z|X,S}(z|x,s)$. If there is none,  select $k_j$ arbitrarily, and if there is  more than one, choose the smallest. 
Set $\ell_j$ to be the bin index of $k_j$, \ie such that $k_j\in\Kset(\ell_j)$.

\item
In block $j+1$, prepare $\rho_{A_{j+1}^n}= \bigotimes_{i=1}^n \Fset^{(s_{j+1,i})} ( \theta_G^{x_{j+1,i}(m_{j+1},\ell_j)   } )$ and send the block $A_{j+1}^n$.
\end{enumerate}
Bob receives the systems $B_1^n,\ldots,B_T^n$ in the state 
\begin{align}
\rho_{B^{Tn}}= \bigotimes_{j=1}^T \bigotimes_{i=1}^n \rho_B^{x_{j+1,i}(m_{j+1},\ell_j)} \,.
\end{align}
Observe that this is the same state as in (\ref{eq:rhoBTnSC}) where the channel $\channel^{(s)}$ is replaced by $\Vset^{(s)}$. 
Thus, Bob can decode reliably and satisfy the distortion requirement, provided that
\begin{align}
R<& I(X,Z;B)_\rho -I(X,Z;S)-\eps(\delta) \,.
\end{align}
This concludes the proof of part 1. %

Part 2 also follows from a similar derivation as  in Appendix~\ref{app:mainSC}, except that now, the state of the input system $A_i$  depends on $(m,s^{i-1},s_i)=(x_i,s_i)$. Hence, we choose the system $G_i$ to be classical, with 
$\theta^{x_i}_{G_i}\equiv
\kb{x_i}$, and then we define the channel $\Fset^{s_i}_{G_i\rightarrow A_i}$ as a preparation channel. Specifically, given the knowledge of $x_i=(m,s^{i-1})$ from the state of $G_i$, the channel $\Fset^{s_i}_{G_i\rightarrow A_i}$ prepares the state 
$\rho_{A_i}^{m,s^{i-1},s_i}=\rho_{A_i}^{x_i,s_i}$.
\qed %

\section{Proof of Theorem~\ref{theo:mainNC}}
\label{app:mainNC}
Consider a random-parameter quantum channel $\channel_{SA\rightarrow B}$ with non-causal CSI at the encoder. 
We show that for every $\zeta_0,\eps_0,\delta_0>0$, there exists a $(2^{n(R-\zeta_0)},n,\eps_0,D+\delta_0)$ code for $\channel_{SA\rightarrow B}$ with non-causal CSI, provided that $(R,D)\in \mathcal{R}_{\text{n-c}}(\channel)$. 
 To prove achievability, we use an extension of  the classical binning technique to the quantum setting, and then apply the quantum packing lemma and the classical covering lemma.

Recall that with non-causal CSI, the encoder has access to the entire sequence of parameters $s_1,s_2,\ldots,s_{n}$ a priori.
Let $\{ p_{X|S}(x|s) , \theta_{A}^{x} \}$  be a  given ensemble, and fix a set of POVMs $\{ \Gamma_{B|x}^{\hs} \}$ such that
\begin{align}
 \sum\limits_{s,\hs,x,z} q(s) p_{X|S}(x|s) %
						\trace( \Gamma_{B|x}^{\hs} \channel^{(s)} (\theta_{A}^{x}) ) d(s,\hs) \leq D \,.
\label{eq:ICncDist}
\end{align}
Define the average states
\begin{align}
\rho^x_B=& \sum_{s\in\Sset} q(s) \channel^{(s)} (\rho^x_A) =\sum_{s\in\Sset} q(s)  \sigma_B^{x,s} %
\intertext{
with
}
\sigma_B^{x,s}=& \channel^{(s)} ( \theta_A^{x,s} )
\end{align}
for $x\in\Xset$.

\subsection*{ }
\vspace{-0.75cm}
The code construction, encoding and decoding procedures are described below.

\subsubsection*{Classical Code Construction}
Let $\delta>0$ and $\tR_s>0$. 
Select $2^{n(R+\tR_s)}$ independent sequences $x^n(m,\ell)$, $m\in [1:2^{nR}]$, $\ell\in [1:2^{n\tR_s}]$,
at random according to $\prod_{i=1}^n p_X(x_{i})$.

\subsubsection*{Encoding and Decoding}
To send the message $m$, given the parameter sequence $s^n$, Alice performs the following.
\begin{enumerate}[(i)]
\item
Find an index $\ell\in [1:2^{n \tR_s}]$ such that $ (s^n, x^n(m,\ell))\in \tset(p_{S,X}) $, where
$p_{S,X}(s,x)=q(s)p_{X|S}(x|s) $. If there is none,  select $\ell$ arbitrarily, and if there is more than one such index, choose the smallest.

\item
Transmit $\rho^{m,\ell}_{A^n}= \bigotimes_{i=1}^n \theta_A^{x_{i}(m,\ell) }$ %
\end{enumerate}
Bob receives the systems $B^n$ in the state 
\begin{align}
\rho_{B^{n}}=  \bigotimes_{i=1}^n \rho_B^{x_{i}(m,\ell)}
\label{eq:rhoBTnNC}
\end{align}
 and decodes as follows. 
\begin{enumerate}[(i)]
\item
Decode $(\hm,\hat{\ell})$ by applying a POVM $\{ \Lambda_{m,\ell} \}_{
(m, \ell) \in  [1:2^{nR}] \times [1:2^{n\tR_s}]}$, which will be specified later. %
\item
Reconstruct the parameter sequence by applying the POVM $\Gamma^{\hs_{i}}_{B|x_{i}}$ to the system $B_{i}$ with
$x_{i}\equiv x_{i}(\hm,\hat{\ell})$, for $i\in [1:n]$.
\end{enumerate}

\subsubsection*{Analysis of Probability of Error and Distortion}
By symmetry, we may assume without loss of generality that Alice sends the message $M=1$ using $L$.
Consider the following events,
\begin{align}
\mathscr{E}_1=& \{  (S^n,X^n(1,\ell))\notin \tset(p_{S,X})  \,,\;\text{for all $\ell\in [1:2^{n\tR_s}] $} \} \\
\mathscr{E}_2=& \{  (\hM,\hL)\neq (1,L)  \}\\
\mathscr{E}_3=& \{  d^n(S^n,\hat{S}^n)>D+\frac{1}{2}\delta_0 \} \,.
\end{align}
By the union of events bound, the probability of error is bounded by
\begin{align}
P_{e|m=1}^{(n)}(\rho_{A^{n}},\Lambda_{B^{n}}) \leq& %
 \prob{ \mathscr{E}_1 }%
+\cprob{ \mathscr{E}_2 }{ \mathscr{E}_1^c } %
+\cprob{ \mathscr{E}_3 }{ \mathscr{E}_1^c \cap \mathscr{E}_2^c }
\label{eq:PeBnc}
\end{align}
where the conditioning on $M=1$ is omitted for convenience of notation.
By the classical covering lemma, the first term tends to zero as $n\rightarrow\infty$ for
\begin{align}
\tR_s> I(X;S)+\eps_1(\delta) \,.
\label{eq:B1nc}
\end{align}

To bound the second term, we use the quantum packing lemma. Given $\mathscr{E}_1^c$, we have that $X^n(1,L)\in\Aset^{\delta_1}(p_X)$, with $\delta_1\triangleq \delta|\Sset| |\Zset|$. Now, observe that
\begin{align}
\Pi^{\delta}(\rho_B)  \rho_{B^n}   \Pi^{\delta}(\rho_B) \preceq& 2^{ -n(H(B)_{\rho}-\eps_2(\delta)) } \Pi^{\delta}(\rho_B)
\\
\trace\left[ \Pi^{\delta}(\rho_B|x^n) \rho_{B^n}^{x^n} \right] \geq& 1-\eps_2(\delta) \\
\trace\left[ \Pi^{\delta}(\rho_B|x^n)  \right] \leq& 2^{ n(H(B|X)_{\rho} +\eps_2(\delta))} \\
\trace\left[ \Pi^{\delta}(\rho_B) \rho_{B^n}^{x^n} \right] \geq& 1-\eps_2(\delta) 
\end{align}
for all $x^n\in\Aset^{\delta_1}(p_X)$, by (\ref{eq:rhonProjIneq}), (\ref{eq:UnitTCond}),  (\ref{eq:PidimCond}), and 
(\ref{eq:UnitTCondB}), respectively. %
Thus, by Lemma~\ref{lemm:Qpacking}, there exists a POVM $\Lambda_{m,\ell}$ such that
$%
\cprob{ \mathscr{E}_2 }{ \mathscr{E}_1^c } \leq 2^{ -n( I(X;B)_\rho -(R+\tR_s)-\eps_3(\delta)) } 
$, %
which tends to zero as $n\rightarrow\infty$, provided that 
\begin{align}
R< I(X;B)_\rho -\tR_s-\eps_3(\delta) \,.
\label{eq:B2nc}
\end{align}

Moving to the third sum in the RHS of (\ref{eq:PeBnc}), suppose that $\mathscr{E}_2^c$ occurred, \ie the decoder measured the correct $M$ and $L$. %
Then,  due to the packing lemma inequality (\ref{eq:QpackB}) and  Lemma~\ref{lemm:gentleM} (the gentle measurement lemma), %
the post-measurement state  $\rho'_{B^n}$ is close to the original state  $\rho_{B^n}$ in the sense that
\begin{align}
\frac{1}{2}\norm{\rho'_{B^n}-\rho_{B^n}}_1 \leq 2^{ -n\frac{1}{2}( I(X;B)_\rho -(R+R_s)-\eps_4(\delta)) } \leq \eps_5(\delta)
\end{align}
for sufficiently large $n$ and rates as in (\ref{eq:B2nc}).
Thus, %
the difference between the probability of a measurement outcome $\hs$ when $\rho'_{B^n}$ is measured and the probability when $\rho_{B^n}$ is measured is bounded by $ \eps_5(\delta)$ in absolute value \cite[Lemma 9.11]{Wilde:17b}.

Therefore, given $\mathscr{E}_1^c \cap \mathscr{E}_2^c\cap \mathscr{E}_3^c$, the parameter sequence $S^n$ and the reconstruction $\hat{S}^n$ have a product distribution according to 
\begin{align}
\prob{ S=s,\hat{S}=\hs }= q(s)  \sum_{x,z} p_X(x) p_{Z|X}(z|x) \trace( \Gamma^{\hs}_{B|x,z} \rho_B^s ) \pm \eps_5(\delta) \,.
\end{align}
By (\ref{eq:ICncDist}), the distribution above satisfies $\E d(S,\hat{S})\leq  D$, hence the last term, $\cprob{ \mathscr{E}_3 }{ \mathscr{E}_1^c \cap \mathscr{E}_2^c }$, tends to zero as $n\rightarrow\infty$ by the law of large numbers. It follows by the law of total expectation, %
\begin{align}
\E d^{n}(S^{n},\hS^{n})\leq \prob{\mathscr{E}_1\cup\mathscr{E}_2\cup\mathscr{E}_3} %
d_{\max} + D+\frac{1}{2}\delta_0 \,. %
\end{align}
 Thereby, the asymptotic average distortion is bounded by $(D+\delta_0)$ and  the probability of error tends to zero  as $n\rightarrow\infty$ for rates that satisfy (\ref{eq:B1nc}) and (\ref{eq:B2nc}), which requires 
\begin{align}
R<& I(X;B)_\rho -I(X;S)-\eps_1(\delta)-\eps_3(\delta) \,.
\end{align}
To show that rate-distortion pairs in $\frac{1}{\kappa}\mathcal{R}_{\text{n-c}}(\channel^{\otimes \kappa})$ are achievable as well, one may employ the coding scheme above for the product channel $\channel^{\otimes \kappa}$, where $\kappa$ is arbitrarily large.
This completes the proof of the direct part.

The converse part follows by the same arguments as in the previous proofs, and it is thus omitted.
\qed %

\section{Proof of Theorem~\ref{theo:mainNone}}
\label{app:mainNone}
Consider a random-parameter quantum channel $\channel_{SA\rightarrow B}$ without CSI. 

\subsection*{Part 1}
Given our previous analysis, the proof of part 1 is straightforward. 
Achievability is shown using the coding scheme in the proof of Theorem~\ref{theo:mainNC} in Appendix~\ref{app:mainNC} with the following modifications. 
The random variable $X$ is statistically independent of the random parameter, \ie $p_{X|S}$ is replaced by $p_X$. The input state does not depend on the random parameter, hence $\theta_A^{x,s}$ is replaced by $\theta_A^x$. 
Set $\tR_s\rightarrow 0$. Hence, $\ell\equiv 1$, the encoding stage (i) is no longer necessary, and the error event $\mathscr{E}_1$ can be ignored. Then, by the same considerations as in Appendix~\ref{app:mainNC}, we have that the asymptotic average distortion is bounded by $(D+\delta_0)$ and the probability of error tends to zero as $n\rightarrow\infty$, provided that
\begin{align}
R< I(X;B)_{\rho}-\eps_3(\delta) \,.
\end{align}
To show that rate-distortion pairs in $\frac{1}{\kappa}\mathcal{R}(\channel^{\otimes \kappa})$ are achievable as well, employ this coding scheme for the product channel $\channel^{\otimes \kappa}$. %
The details are omitted.

The converse proof also follows similar arguments as in the previous proofs.  Suppose that Alice  sends 
$\rho^{m}_{ A^n}$ over the channel. 
Bob receives the systems $B^n$ in the state
$\rho_{B^n}=\frac{1}{2^{nR}} \sum_{m=1}^{2^{nR}} \sum_{s^n\in\Sset^n} q^n(s^n) \rho_{B^n}^{m,s^n}$, with  
\begin{align}
\rho_{B^n}^{m,s^n}\equiv \channel^{(s^n)}(\rho^{m}_{ A^n}) 
\end{align}
and then, performs a decoding POVM $\Lambda^{m,\hs^n}_{B^n}$. 
Now,
consider a sequence of codes $(\Eset_{M\rightarrow A^n},$ $\Lambda_{B^n})$ such that the average probability of error tends to zero and the distortion requirement holds. That is,
\begin{align}
&P_e^{(n)} (\Eset,\Lambda) \leq \alpha_n \,,
\label{eq:convPe1None}
\intertext{and}
& \Delta^{(n)} (\Eset,\Lambda)  \leq D \,.
\end{align}
By Fano's inequality, (\ref{eq:convPe1None}) implies that
$%
H(M|\hM) \leq n\eps_n
$, %
where $\eps_n$ tends to zero as $n\rightarrow\infty$.
Hence, 
\begin{align}
nR= H(M) \leq& I(M;\hM)_{\rho}+n\eps_n %
\leq I(M;B^n)_{\rho}+n\eps_n 
\label{eq:ConvIneq1none}
\end{align}
where the last inequality follows from the Holevo bound \cite[Theorem 12.1]{NielsenChuang:02b}.
Thus,
\begin{align}
R\leq \frac{1}{n} I(M;B^n)_{\rho}+ \eps_n = \frac{1}{n}  I(X^n;B^n)_{\rho}+ \eps_n
\end{align}
for $X^n=f(M)$  where $f$ is an arbitrary one-to-one map from $[1:2^{nR}]$ to $\Xset^n$.

As for the distortion requirement,
\begin{align}
D&\geq
 \Delta^{(n)}(\Eset,\Lambda)= \E d^n(S^n,\hS^n) \nonumber\\&= 
P_e^{(n)} (\rho_{A^n},\Lambda_{B^n}) \E[ d^n(S^n,\hS^n) \,|\; \hM\neq M ]\nonumber\\& + (1- P_e^{(n)} (\rho_{A^n},\Lambda_{B^n})) \E[ d^n(S^n,\hS^n) \,|\; \hM= M ]
\nonumber\\
&\geq  (1- P_e^{(n)} (\rho_{A^n},\Lambda_{B^n})) \E[ d^n(S^n,\hS^n) \,|\; \hM= M ]
\nonumber\\
&\geq (1-\alpha_n)  \E[ d^n(S^n,\hS^n) \,|\; \hM= M ]
\end{align}
where we have used the law of total expectation in the second line, and (\ref{eq:convPe1None}) in the last line. 
Thus,  
\begin{align}
D&\geq   \E[ d^n(S^n,\hS^n) \,|\; \hM= M ] -\alpha_n d_{\max} 
\nonumber\\
&= \sum_{s^n\in\Sset^n} \sum_{\hs^n\in\widehat{\Sset}^n}
d^n(s^n,\hs^n) q^n(s^n) \cdot \frac{1}{2^{nR}} \sum_{m=1}^{2^{nR}} \trace\Big[ \Lambda^{m,\hat{s}^n}_{B^n}
\channel^{(s^n)}_{A^n\rightarrow B^n}
(\overline{\rho}_{A^n}^{m}) \Big] -\alpha_n d_{\max}
\label{eq:distortionNoneconvN}  \\ 
&= \sum_{s^n\in\Sset^n}  \sum_{x^n\in\Xset^n} %
\sum_{\hs^n\in\widehat{S}^n} q^n(s^n) p_{X^n}(x^n)%
 \trace(  \Gamma_{B^n|x^n}^{\hs^n} \rho^{s^n,x^n}_{B^n} ) d^n(s^n,\hs^n)
\end{align}
with $\Gamma_{B^n|f(m)}^{\hs^n}\equiv \Lambda^{m,\hat{s}^n}_{B^n}$.

The union in (\ref{eq:inCeaCnc}) is restricted to pure states $ \theta^{x}_{A}= \kb{ \phi^{x}_{A} } $ with
$|\Xset|\leq |\Hset_A|^2+1$, based on the 
same arguments as in the proof of Lemma~\ref{lemm:pureCeaSCausal} in Appendix~\ref{app:pureCeaSCausal}. 
This concludes the converse proof for part 1.

\subsection*{Part 2}
Now, we consider an entanglement-breaking channel. The direct part follows from 
part 1, taking $\kappa=1$.
It remains to prove the converse part, which we show by extending the methods of Wang \etal
\cite{WangDasWilde:17p}.
By (\ref{eq:ConvIneq1none}), we have
\begin{align}
R&\leq \frac{1}{n} I(M;B^n)_{\rho}+\eps_n 
\nonumber\\&
= \frac{1}{n}\sum_{i=1}^n I(M;B_i|B^{i-1})_{\rho}
\nonumber\\&
\leq \frac{1}{n}\sum_{i=1}^n I(M,B^{i-1};B_i)_{\rho}
\label{eq:NoSIconv1}
\end{align}
by the chain rule.
Without CSI, the channel input systems $A^n$ have no correlation with the channel parameter sequence $S^n$. As the channel is memoryless, it follows that $B_i$ and $S^{i-1}$ are in a product state. Then,
\begin{align}
I(M,B^{i-1};B_i)_{\rho}&\leq I(M,B^{i-1},S^{i-1};B_i)_{\rho} \nonumber\\&
=I(M,B^{i-1};B_i|S^{i-1})_{\rho}+I(S^{i-1};B_i)_{\rho} \nonumber\\&
=I(M,B^{i-1};B_i|S^{i-1})_{\rho}  
\label{eq:NoSIconv2}
\end{align}
where the last equality holds since $I(S^{i-1};B_i)_{\rho}=0$.

If the random-parameter quantum channel  is  entanglement breaking, then $\channel_{A\rightarrow B}^{(s)}$ can be presented as a concatenation of a measurement channel, followed by a state-preparation channel, i.e.
\begin{align*}
\channel^{(s)}_{A\rightarrow B}= \Pset^{(s)}_{Y_s\rightarrow B}\circ \Mset^{(s)}_{A\rightarrow Y_s}
\end{align*}
where $Y_s$ is classical, for $s\in\Sset$ (see Subsection~\ref{subsec:Qchannel}).
Therefore, by the quantum data processing theorem due to Schumacher and Nielsen \cite{SchumacherNielsen:96p}\cite[Theorem 11.9.4]{Wilde:17b}, 
\begin{align}
I(M,B^{i-1};B_i|S^{i-1}=s^{i-1})_{\rho} \leq I(M,Y_{s^{i-1}}^{i-1};B_i|S^{i-1}=s^{i-1})_\rho \,.
\label{eq:NoSIconv3}
\end{align} 

By (\ref{eq:NoSIconv1})-%
(\ref{eq:NoSIconv3}), we have
\begin{align}
R-\eps_n&\leq \frac{1}{n} \sum_{i=1}^n I(M,Y_{S^{i-1}}^{i-1};B_i|S^{i-1})_\rho \nonumber\\&
\leq \frac{1}{n} \sum_{i=1}^n I(X_i;B_i)_\rho %
\label{eq:NoSIconv4}
\end{align}
with $X_i\equiv (M,Y_{S^{i-1}}^{i-1},S^{i-1})$.
 Define
\begin{align}
X\equiv X_J \,,\; S \equiv S_J \,,\; \hS\equiv \hS_J
\label{eq:convSCxsyhsNoneEB}
\end{align}
where $J$ is a classical time-sharing variable that is uniformly distributed over $[1:n]$. Observe that for this choice of $X$,
we have
\begin{align}
\rho_{XB}\equiv \sum_{x} p_{X}(x) \kb{ x} \otimes \channel(\phi_{A}^x)
=\frac{1}{n} \sum_{i=1}^n \sum_{x_i} p_{X_i}(x_i) \kb{ x_i} \otimes \channel(\sigma_{A_i}^{x_i}) %
\end{align}
where $\sigma_{A_i}^{m,y^{i-1}_{s^{i-1}},s^{i-1}}=\rho_{A_i}^m$.
Thus, by (\ref{eq:NoSIconv4}),
\begin{align}
R\leq&  I(X;B|J)_\rho +\eps_n\leq I(X';B) +\eps_n
\end{align}
with $X'\equiv (X,J)$.

As for the distortion, 
\begin{align}
D\geq& \E d^n(S^n,\hS^n)= \frac{1}{n} \sum_{i=1}^n \E d(S_i,\hS_i) =\E d(S,\hS)   
\end{align}
where the first equality holds since the distortion measure is additive (see (\ref{eq:dn})), and the second follows from the definition of $S$ and $\hS$ in (\ref{eq:convSCxsyhsNoneEB}).
This completes the proof of Theorem~\ref{theo:mainNone}.
\qed

\end{appendices}

\bibliography{references2}{}

\end{document}